\documentclass[10pt,letterpaper,twocolumn]{article}
\usepackage[margin=0.65in,top=0.72in,bottom=0.72in]{geometry}
\usepackage[T1]{fontenc}
\usepackage{lmodern}
\usepackage{cite}
\usepackage{amsmath,amssymb,amsthm}
\usepackage{algorithm,algorithmic}
\usepackage{graphicx}
\usepackage{textcomp}
\usepackage{booktabs,array}
\usepackage[font=small,labelfont=bf]{caption}
\usepackage{xurl}
\usepackage[hidelinks]{hyperref}
\graphicspath{{figures/}}
\newcommand{\qnEoneRatioFine}{936--1{,}105}  %
\newcommand{\qnEoneRatioGrid}{1{,}037--1{,}364}  %
\newcommand{\qnLbOverMeasUniform}{521--695}  %
\newcommand{\qnVaultNninetyRtenLow}{2{,}160}  %
\newcommand{\qnVaultNninetyRtenHigh}{126}  %
\newcommand{\qnVaultConst}{4.2--5.6}  %
\newcommand{\qnCrossDep}{0.22--0.30}  %
\newcommand{\qnCrossDeph}{0.18--0.25}  %
\newcommand{\qnProvedUniform}{134--143}  %
\newcommand{\qnProvedUniformMin}{134}  %
\newcommand{\qnProvedWr}{13.1--13.8}  %
\newcommand{\qnProvedWrThm}{13.8--14.6}  %
\newcommand{\qnProvedWhalf}{24.6--26.1}  %
\newcommand{\qnLbOverMeasWr}{53--55}  %
\newcommand{\qnLbOverMeasWhalf}{86--111}  %
\newcommand{\qnAdaptiveOverLb}{1.36--2.12}  %
\newcommand{\qnWhalfMeas}{219--221}  %
\newcommand{\qnWrMeas}{75--96}  %
\newcommand{\qnWrSlope}{0.45--0.58}  %
\newcommand{\qnEsevenRatio}{231--319}  %
\newcommand{\qnLocalRatio}{2.4--3.8}  %
\newcommand{\qnInsealObl}{1.0--1.2}  %
\newcommand{\qnExtendGivenR}{0.84--1.00}  %
\newcommand{\qnSparseRatio}{1.9--2.3}  %
\newcommand{\qnMirrorVault}{0.80--1.19}  %
\newcommand{\qnMirrorInseal}{0/100}  %
\newcommand{\qnMirrorPaired}{2/50}  %
\newcommand{\qnExtendInseal}{0/100}  %
\newcommand{\qnOracleNotInC}{0/1{,}200}  %
\newcommand{\qnOracleInCRatio}{0.71--1.20}  %
\newcommand{\qnProductRatio}{0.79--1.59}  %
\newcommand{\qnProductRatioInterp}{0.91--1.45}  %
\newcommand{\qnProductSuccess}{1{,}578/1{,}600}  %
\newcommand{\qnVaultBoundRtenLow}{14{,}923}  %
\newcommand{\qnVaultBoundRtenHigh}{933}  %
\newcommand{\qnPsignOverVaultFineLow}{76--169}  %
\newcommand{\qnPsignOverVaultGridLow}{64--158}  %
\newcommand{\qnPsignOverVaultFineHigh}{5.6--9.0}  %
\newcommand{\qnPsignOverVaultGridHigh}{4.7--8.2}  %
\newcommand{\qnPsignSignCost}{1.00--1.20}  %
\newcommand{\qnPsignSuccess}{560/560}  %
\newcommand{\qnPsignBoundLow}{0.61--1.30}  %
\newcommand{\qnPsignBoundHigh}{2.4--5.1}  %
\newcommand{\qnProvedPsignRten}{12.8}  %
\newcommand{\qnProvedPsignReight}{3.1}  %
\newcommand{\qnProvedExtendPerUnit}{70--75}  %
\newcommand{\qnCrossR}{5}  %
\newcommand{\qnTierThreeVaultDetect}{0.97}  %
\newcommand{\qnTierThreeVaultFA}{0/300}  %
\newcommand{\qnTierThreeBlind}{0.00}  %
\newcommand{\qnTierThreeRegistry}{1.00}  %
\newcommand{\qnTierThreeTimeErr}{1.2--2.9}  %
\newcommand{\qnCircuitChecks}{29}  %
\newcommand{\qnCircuitLawChecks}{27}  %
\newcommand{\qnCircuitMaxDev}{6\times10^{-16}}  %
\newcommand{\qnCircuitNoFrame}{0.125}  %
\newcommand{\qnCircuitFrameAvg}{4\times10^{-16}}  %
\newcommand{\qnResourceRows}{4 & 0.15 & 989 & 31{,}648 & 13{,}457 & 14 \\
4 & 0.6 & 49 & 1{,}568 & 885 & 18 \\
6 & 0.15 & 1{,}456 & 46{,}592 & 88{,}948 & 61 \\
6 & 0.6 & 70 & 2{,}240 & 4{,}136 & 59 \\
8 & 0.15 & 1{,}626 & 52{,}032 & 440{,}015 & 271 \\
8 & 0.6 & 98 & 3{,}136 & 20{,}581 & 210 \\
10 & 0.15 & 2{,}160 & 69{,}120 & 2{,}178{,}915 & 1{,}009 \\
10 & 0.6 & 126 & 4{,}032 & 117{,}881 & 936 \\}  %
\newcommand{\qnVaultQubitsRtenLow}{69{,}120}  %
\newcommand{\qnVaultQubitsRtenHigh}{4{,}032}  %
\newcommand{\qnLifeVaultDep}{0.310}  %
\newcommand{\qnLifeVaultDeph}{0.356}  %
\newcommand{\qnLifeSeal}{0.0066}  %
\newcommand{\qnChenLbRten}{3{,}848}  %
\newcommand{\qnChenVaultUbRten}{7{,}834}  %
\newcommand{\qnChenRstar}{11}  %
\newcommand{\qnSealRhoTwo}{1.3--1.4}  %
\newcommand{\qnSealRhoFive}{2.5--2.8}  %
\newcommand{\qnSealLifeFive}{0.0689}  %
\newcommand{\qnSealRhoTen}{9.3--10.2}  %
\newcommand{\qnSealLifeTen}{0.1431}  %
\newcommand{\qnSealLifeGainFive}{10}  %
\newcommand{\qnSealRatioChange}{0.77--1.03}  %
\newcommand{\qnSealPOne}{35/36}  %
\newcommand{\qnSealPTwo}{12/12}  %
\newcommand{\qnSealPThree}{10/10}  %
\newcommand{\qnSealFailRho}{0.94 [0.86, 1.03]}  %
\newcommand{\qnSealFailBand}{[1.06, 1.08]}  %
\newcommand{\qnReorderFstar}{5\%}  %
\newcommand{\qnReorderDetectFive}{0.93--1.00}  %
\newcommand{\qnReorderFailTwenty}{4}  %
\newcommand{\qnReorderFATwenty}{74/300}  %
\newcommand{\qnReorderDetectFull}{0.23--0.51}  %
\newcommand{\qnReorderFATotal}{999}  %
\newcommand{\qnCaseUnits}{648}  %
\newcommand{\qnCaseWindow}{166}  %
\newcommand{\qnCaseOnset}{56}  %
\newcommand{\qnCaseSwitchTrue}{321}  %
\newcommand{\qnCaseSwitchFound}{322}  %
\newcommand{\qnCaseToolOne}{$-$\texttt{ZIIZZY}}  %
\newcommand{\qnCaseToolTwo}{$-$\texttt{ZIIZYZ}}  %

\newtheorem{proposition}{Proposition}
\newtheorem{lemma}{Lemma}
\theoremstyle{definition}
\newtheorem{definition}{Definition}

\newcolumntype{L}[1]{>{\raggedright\arraybackslash}p{#1}}
\newcommand{\eps}{\varepsilon}
\newcommand{\astar}{a^{*}}
\newcommand{\Nn}{N_{90}}
\newcommand{\Reg}{R}
\newcommand{\ceff}{c^{*}}

\newcommand{\authorbio}[3]{%
  \par\medskip\noindent
  \begin{minipage}[t]{0.78in}
    \vspace{0pt}#1
  \end{minipage}\hfill
  \begin{minipage}[t]{\dimexpr\columnwidth-0.90in\relax}
    \small\textbf{#2} #3
  \end{minipage}\par
}

\title{Quantum-Memory Forensics for Tampered Quantum Seals: Resource-Dependent Sample Gains}
\author{Jaeyun Jeong\textsuperscript{1}, Zehua Wang\textsuperscript{2}, Lilia Tightiz\textsuperscript{3},
  Ki-Woong Park\textsuperscript{1}, Victor C. M. Leung\textsuperscript{2}}
\date{}
\hypersetup{
  pdftitle={Quantum-Memory Forensics for Tampered Quantum Seals: Resource-Dependent Sample Gains},
  pdfauthor={Jaeyun Jeong, Zehua Wang, Lilia Tightiz, Ki-Woong Park, Victor C. M. Leung}
}

\begin{document}
\twocolumn[{
\begin{@twocolumnfalse}
\maketitle
\vspace{-1.8em}
\begin{center}
\small
\textsuperscript{1} Department of Computer and Information Security, Sejong University, Seoul 05006, Republic of Korea\\
\textsuperscript{2} Department of Electrical and Computer Engineering, The University of British Columbia, Vancouver, BC V6T 1Z4, Canada\\
\textsuperscript{3} Department of Computer Science and Engineering, Sejong University, Seoul 05006, Republic of Korea\\[2pt]
\texttt{3500jjy@gmail.com}, \texttt{zwang@ece.ubc.ca}, \texttt{liliatightiz@sejong.ac.kr},\\
\texttt{woongbak@sejong.ac.kr}, \texttt{vleung@ece.ubc.ca}\\[2pt]
Correspondence to Ki-Woong Park (\href{mailto:woongbak@sejong.ac.kr}{\texttt{woongbak@sejong.ac.kr}}).
\end{center}
\begin{abstract}
Quantum seals reveal that an object was disturbed, but not where, with which tool, or when the tool changed. We study
quantum-memory forensics for sealed items returning through an untrusted network where one tool opens many. Each seal cell is half of a Bell pair whose partner the verifier keeps in a protected quantum memory,
a secret Pauli frame turns any attack into a Pauli channel, and a three-tier report localizes the tampering, learns a
hidden-parity fingerprint of the tool and detects a tool switch in arrival order, not necessarily in attack order.
Attribution concerns attackers who do not erase that
fingerprint, which a random Pauli layer would do cheaply. Each result is stated within one of four verifier classes.
For per-unit verifiers, which never couple two seal units and hold no untouched
reference, the sample separation is asymptotically exponential, by reduction from a known lower bound for one parity
of strength up to 1/3. On ten tampered cells it is proved to be at least \qnProvedUniformMin-fold against non-adaptive
stabilizer probes and observed in simulation at \qnEoneRatioFine-fold. Against an implemented self-referencing
strategy the simulated gain is about twice the inverse squared strength. The value also depends on the attack: destroying the
untouched cells makes the reference necessary, whereas copying errors exposes the fingerprint to reference-free
probes. We derive the vault's qubit inventory and simulate a seal requirement curve: ten times more seal noise
lengthens the seal's lifetime \qnSealLifeGainFive-fold at \qnSealRhoFive{} times the attribution cost at $r=10$.
\end{abstract}

\begin{quote}
\end{quote}

\begin{quote}
\small
This work has been submitted to the IEEE for possible publication. Copyright may be transferred without notice,
after which this version may no longer be accessible.
\end{quote}
\vspace{0.75em}
\end{@twocolumnfalse}
}]

\section{Introduction}
\label{sec:introduction}

A quantum seal protects an object with quantum states whose disturbance reveals tampering
\cite{williams2016tamper}, and tamper-detection codes answer the same question for quantum data: was it disturbed?
\cite{tamperunitary2023,universaltamper2025}. A security investigation needs more than that answer. Consider an
operator who ships seal modules to a supplier, receives the sealed items back through a logistics network it does not
control, and checks each item on arrival. If many items arrive tampered, the operator needs to know whether one tool
opened them all, which seal cells it reached, from which item on, and whether the attacker changed tools along the way.
These answers separate one compromised hub from scattered incidents and link an incident to earlier ones. Each
triggers a different response: a single tool points to one hub to audit rather than the whole network, the onset
separates the items that can be released from those that must be recalled, and a change of tool shows that the
attacker has adapted. Detection alone, which every tamper-evident seal provides, triggers none of these. We call
these questions \emph{attribution}, and the procedure that answers them \emph{forensics}. They are hardest when the
attacker is quantum-capable, because such an attacker can scramble the disturbance: rate tests still detect the
attack, but the structure that identifies the tool is hidden.

This paper studies quantum-memory forensics (QMF). Each seal cell is one half of a Bell pair, and the verifier keeps
the other half in a protected quantum memory, the \emph{vault}. A secret random Pauli frame, applied at issue and
undone at verification, twirls any attack into a Pauli channel \cite{bennett1996mixed,dankert2009exact}, and a Bell
measurement of a cell against its vault partner yields one sample of the attack's Pauli error per seal unit. We model
the tool by a hidden-parity fingerprint: a Pauli $\astar$ on the tampered region whose parity the tool leaves biased
with strength $\eps$ and sign $s$. Physically, $\astar$ is a quantity conserved by the tool's dynamics and imperfectly
hidden by the attacker. From the vault's records QMF returns a three-tier report: the tampered cells, their error rates
and the onset of tampering; the fingerprint $(\astar,s,\eps)$; and the time of a switch between tools, with links to
earlier incidents. Attribution is up to fingerprint equivalence: tools that leave the same fingerprint on the same
cells are not told apart, and a link to an earlier incident means the same fingerprint, not the same device.

The fingerprint is exactly the object of Pauli channel learning, where learning with a quantum memory is known to be
exponentially more sample-efficient than learning without one
\cite{chen2022quantum,chen2024tight,chen2021exponential,huang2022quantum}. It is tempting to conclude that a vault gives
forensics an exponential advantage. We found that this conclusion holds only for some verifiers and some attacks, for
three reasons. Untouched cells of the same seal are themselves a reference: against an attacker who leaves them alone,
a verifier that pairs tampered with untouched cells does as well as the vault. An attacker who destroys those cells
removes that reference and makes the vault necessary for every verifier that probes one unit at a time. And an
attacker who copies errors onto other cells blinds Bell-pair references but hands the fingerprint to verifiers with no
reference at all, because two copies can be read in complementary bases. Two seal units entangled with each other can
also stand in for a memory, if the deployment lets the verifier measure them together.

We therefore fix the operating conditions first and state every result within one of four verifier classes: per-unit
verifiers, which never couple two seal units and probe only cells within the attacker's reach; self-referencing
verifiers, which entangle two units; in-seal reference verifiers, which use untouched cells; and the protected
reference, the vault. The sample gain of a protected reference then depends on the allowed resources and on the
attack structure. In the specified per-unit resource model an asymptotically exponential separation holds. Against
the implemented
self-referencing alternative the difference is a simulated polynomial factor, and we claim no optimality over the
whole self-referencing class.

What is new relative to Pauli channel learning is the task and its setting, not the learning separation, which we
inherit. Channel learning estimates the parameters of a channel to which the learner has clean, chosen access.
Forensics starts from an uncontrolled record: it must find which cells were touched and from which unit on, decide
whether a fingerprint is present, attribute it, detect a change of tool, and say when it cannot
(Section~\ref{sec:report}). Its resources are set by the deployment and the attacker rather than by the
experimenter: whether untouched cells survive, whether two items can be held together, and what the attacker does to
the cells it does not need. Along these axes the separation appears or vanishes (Section~\ref{sec:res-attack}), and
the learning bounds, which take the channel itself as the object of study, do not address them.

\subsection{Contributions}
The paper makes four contributions, each stated with its scope. The main one is a map of when a protected reference
pays off, by verifier class and by attack. Against per-unit verifiers the gain is asymptotically exponential when
untouched references are absent or destroyed, by reduction from the lower bound of Chen \emph{et al.} for one parity
with $\eps\le1/3$ \cite{chen2024tight}. At ten tampered cells it is proved to be at least \qnProvedUniform{} against
non-adaptive stabilizer probes and observed in simulation at \qnEoneRatioFine. Against the implemented
self-referencing strategy the gain is a simulated factor of about $2/\eps^2$. Against in-seal references facing an
oblivious attacker there is none, and copying attackers expose the fingerprint to reference-free product probes
(Section~\ref{sec:results}).

The second contribution is a forensic report with a stated scope. QMF localizes tampering, decodes the fingerprint
and detects a tool switch. It returns explicit status codes when an attack falls outside its supported scope, which
is an oblivious tool on fewer than half of the cells, one switch between tools with different fingerprints, and a
record in which the tampered units follow the onset contiguously (Section~\ref{sec:report}). Tool-switch detection is
evaluated in simulation in that scope; incident linking is only demonstrated.
Detection and localization (tier~1) and attribution (tiers~2 and~3) are separate goals, and
Section~\ref{sec:claim-scope} states what the report delivers against each attacker. Under extend, spray and mirror the
fingerprint is recovered only by conditional decoders; the full report flags these attacks with a status code but
does not decode them. Third, we turn the results into engineering requirements for one deployment: the number of tampered items needed for
attribution, the vault's qubit inventory, and the coherence that vault and seal need. In the simulated range the vault
tolerates memory noise up to 0.2 (depolarizing). Under independent noise the seal's lifetime, not the vault's, is the
binding constraint, and the seal's noise budget is a
simulated trade-off: ten times more seal noise lengthens the seal's lifetime \qnSealLifeGainFive-fold at \qnSealRhoFive{}
times the attribution cost (Section~\ref{sec:engineering}). Fourth, a Fano bound in terms of the largest coverage of a weight-$w$ Pauli by one
stabilizer group turns the simulated gaps into ratios of a lower bound to an upper bound (Section~\ref{sec:theory}).
For full weight the coverage is a known theorem \cite{tran2016correlations,eltschka2020maximum}, which our
enumerations confirm independently. Section~\ref{sec:limitations} lists where the vault gives no exponential gain:
fingerprints of weight at most three, unscrambled tampering, known fingerprints, copying attackers and attackers who
erase the fingerprint with a random Pauli layer, which is cheap; forensics concerns attackers who do not.

The protocols and decision rules of the experiments were written in a laboratory notebook, with post-hoc analyses
marked; a register sets each notebook time against the logged start of the run and states where the order is not
documented. Every result figure and table is rebuilt from hashed raw results by the released code (Section~\ref{sec:setup}). Section~\ref{sec:related} reviews related work. Sections~\ref{sec:model}
and~\ref{sec:report} define the scenario, the model, the verifier classes and the QMF report.
Section~\ref{sec:theory} gives the bounds and Section~\ref{sec:setup} the evaluation protocol.
Section~\ref{sec:results} reports the results, Section~\ref{sec:engineering} turns them into requirements,
Section~\ref{sec:limitations} lists the boundaries and limitations, and Section~\ref{sec:conclusion} concludes.
Appendices~A--C give the proofs, the physical realization of the tamper model and the statistics; the supplement
gives the experiment register and further results.

\section{Related Work}
\label{sec:related}

\subsection{Quantum seals and tamper detection}
Quantum seals use the disturbance of quantum states to reveal access to an object \cite{williams2016tamper}.
Tamper-detection codes and related constructions detect, with high probability, whether a quantum message or state
was modified \cite{tamperunitary2023,splitstate2023,tamperevident2024,universaltamper2025}. These works decide accept
or reject. None of them attributes the tampering, that is, where it happened, with which tool and when the tool
changed, and none uses a verifier-held reference to learn the attack's structure. QMF keeps detection as tier~1 and
adds attribution in tiers~2 and~3.

\subsection{Learning with and without quantum memory}
A line of work shows that access to a quantum memory can reduce the number of experiments needed to learn properties
of states and channels exponentially \cite{chen2021exponential,huang2021information,aharonov2022quantum}, and Huang
\emph{et al.} demonstrated such gains on hardware \cite{huang2022quantum}. For Pauli channels, estimating the Pauli
eigenvalues with an entangled ancilla takes $O(n/\eps^2)$ channel uses \cite{chen2022quantum}, whereas without
entanglement it takes $\Omega(2^n/\eps^2)$ even with adaptivity and classical memory \cite{chen2024tight}. The hard
instance of the latter is a hidden-parity family, which is the fingerprint model we use. Error rates and Pauli error
probabilities, in contrast, can be estimated efficiently without entanglement
\cite{flammia2020efficient,harper2020efficient,harper2021fast,flammia2021pauli}, which is why tier~1 shows no
separation. Kim and Oh separate the two resources that a quantum memory supplies: an exponential advantage needs
enough ancilla qubits, while the amount of entanglement can be small, and without ancillas even the low-weight Pauli
eigenvalues need a number of channel uses exponential in the weight \cite{kim2026fundamental}. Their lower bound, like
that of Chen \emph{et al.}, uses channels with a single non-trivial eigenvalue, our fingerprint model. In these terms a
per-unit verifier has no ancilla outside the attacker's reach, while the untouched cells of an in-seal reference and
the vault supply one. We inherit the separation and do not claim it as new. Those results fix the learner's
resources and hand it the channel; in forensics the resources are set by the deployment and the attacker, the record
must first be localized, and the output must say when attribution is impossible. Our questions are therefore which
verifiers the separation applies to in a forensic deployment, how attacks change it, and what a vault and a seal must
deliver in practice.

\subsection{Stabilizer coverage and sector lengths}
A non-adaptive stabilizer probe learns the parity of every Pauli in its stabilizer group. How much of a weight-$w$
family a single group can cover therefore bounds what such probes can learn. For full weight, the largest number of
weight-$r$ elements is the maximal $r$-body sector length of an $r$-qubit state. That maximum is
$2^{r-1}+[r\ \text{even}]$ and is attained by the GHZ state, as shown by Tran \emph{et al.} for odd $r$
\cite{tran2016correlations} and by Eltschka and Siewert for even $r$ \cite{eltschka2020maximum}. We use this theorem in
our Fano bound and confirm it by enumeration; we do not claim it.

\subsection{Device fingerprints}
Noise fingerprints have been used to identify honest quantum devices from their outputs
\cite{martina2022learning,priestley2026fingerprint}. QMF instead attributes an \emph{attack}: the fingerprint belongs
to the tampering tool, and the verifier holds a reference that the attacker cannot reach.

\section{System and Threat Model}
\label{sec:model}

\begin{figure*}[t]
\centering
\includegraphics[width=\textwidth]{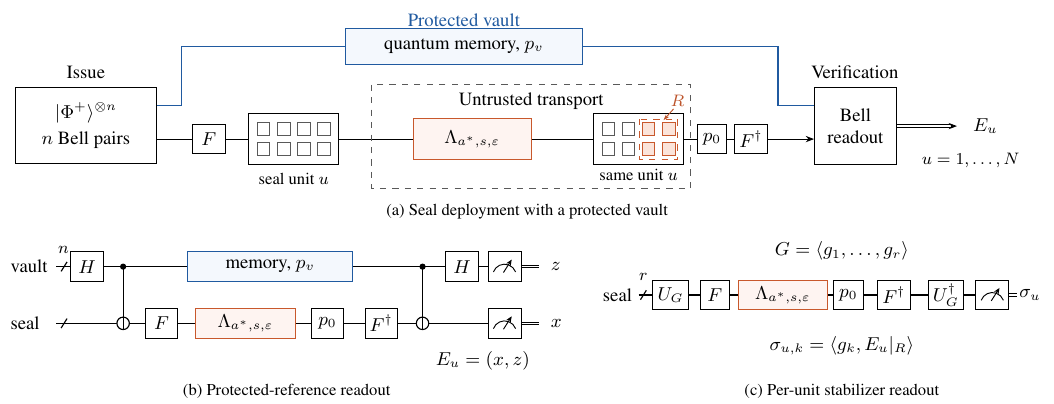}
\caption{QMF system model and measurement resources.
(a)~Each of the $n$ cells of a seal unit is paired with a qubit retained in the protected vault.
The same unit is shown before and after untrusted transport; orange cells mark the attacked region $\Reg$ of
$r$ cells. Cell arrays are schematic. The tool $\Lambda_{\astar,s,\eps}$ acts on $\Reg$; $p_0$ and $p_v$ denote
benign seal and vault noise. The secret frame $F$ is applied at issue and undone at verification.
(b)~Vault circuit, with $n$ cells per bundled wire. Bell measurement returns a full Pauli error sample
$E_u=(x,z)\in\{I,X,Y,Z\}^{n}$ for each arriving unit.
(c)~A per-unit stabilizer probe on $r$ reachable cells returns $r$ syndromes of a chosen group $G$,
without an untouched reference.}
\label{fig:structure}
\end{figure*}

\subsection{Operating scenario}
\label{sec:scenario}
We fix one deployment and define every quantity of the model in it. Fig.~\ref{fig:structure}(a) shows this
deployment; panels (b) and (c) show the vault's readout and a per-unit probe. A verifier, for example the receiving
inspection site of an operator, prepares Bell pairs. It sends one half of each pair, as an $n$-cell seal module, to a
supplier and keeps the other half in its vault. The supplier fits one module to each item of a consignment and ships
the items back through a logistics network that the verifier does not control. At one hub of that network an
attacker opens many items with the same tool and reseals them. The items reach the inspection site in an order the
verifier does not choose, and each item is verified and released on arrival.

The \emph{unit} of the analysis is one seal module on one item. The attacker's tool is the same physical device each
time, so the same tamper channel acts on every unit it touches. Attribution then asks how many tampered units the
verifier must see before it can name the tool. Its cost $\Nn$ (Section~\ref{sec:task}) is therefore an
\emph{attribution threshold}: the number of items the attacker must touch with one tool before the verifier
identifies the tool with probability $0.9$. A single disturbance of a single unit cannot be attributed by any
verifier.

The vault holds one qubit per cell of every module that has been issued and not yet verified. When a consignment of
$N$ modules is issued at once, the vault holds $nN$ qubits for the round-trip time $T$ of the modules. Each of those
qubits, and each seal cell, must stay coherent for $T$. Section~\ref{sec:engineering} turns these requirements into
numbers.

\subsection{Seal, vault and frame}
At issue, cell $c$ of unit $u$ is prepared as one half of $|\Phi^+\rangle$, the other half is stored in the vault, and
a uniformly random Pauli $F_{u,c}$, the frame, is applied to the cell and recorded as a classical secret. At
verification the verifier undoes the frame and measures each cell jointly with its vault partner in the Bell basis.
The outcome is a Pauli $E_u\in\{I,X,Y,Z\}^{n}$, one sample of the error the cell suffered between issue and
verification. Averaging over the secret frame twirls an arbitrary attack into a Pauli channel
\cite{bennett1996mixed,dankert2009exact} (Appendix~B derives this), so the Bell measurement samples the attack's Pauli error distribution,
composed with the benign noise. Exact simulation of a coherent attack, RX$(0.7)\otimes$RX$(0.4)$ on a GHZ probe, shows
what the frame does: without it the outcome law deviates from the Pauli-channel prediction by \qnCircuitNoFrame;
averaged over the 16 frames it matches to within $\qnCircuitFrameAvg$.

\subsection{Tamper model and its physical motivation}
\label{sec:tamper}
The tool acts on a region $\Reg$ of $r$ cells. We model it by the hidden-parity channel
\begin{equation}
p_{\Reg}(E)=4^{-r}\left(1+s\,\eps\,(-1)^{\langle \astar,E\rangle}\right),
\label{eq:scramble}
\end{equation}
where $\langle\cdot,\cdot\rangle$ is the symplectic product, $\astar\neq I$ is the fingerprint, $s\in\{\pm1\}$ its
sign and $\eps\in(0,1]$ its strength. Its Pauli eigenvalues are $\lambda_I=1$, $\lambda_{\astar}=s\eps$ and $0$
otherwise. The weight $w=|\astar|$ is the number of cells on which $\astar$ acts. For $w\ge2$ every cell of $\Reg$,
and every set of cells that does not contain the support of $\astar$, is fully depolarized, so only the joint parity
of $\astar$ carries the fingerprint. For $w=1$ the one cell in the support keeps a single-qubit bias $s\eps$, and its
error rate differs from $3/4$ by $s\eps/4$; such fingerprints are $3r$ of the $4^r-1$ in the uniform family.

Why should a scrambling tool leave such a parity? We read (\ref{eq:scramble}) as a \emph{quantity conserved by the
tool's dynamics} that the attacker has hidden imperfectly. Suppose the tool's internal operations commute with a
Pauli $\astar$, a symmetry of its coupling to the cells. Those operations can randomize every Pauli eigenvalue except
$\lambda_{\astar}$, which they leave at $\pm1$. An attacker who knows this adds symmetry-breaking kicks, a Pauli that
anticommutes with $\astar$, applied with probability $p$. The residual eigenvalue is $1-2p$, so $\eps=|1-2p|$ and
$s=\mathrm{sgn}(1-2p)$. Hiding the parity by calibration needs $p=1/2$ exactly, and any miscalibration leaves
$\eps>0$. \emph{Erasing} it, however, is cheap. An attacker with coherent access to the cells who applies, after the
tool, an independent uniformly random Pauli $P$ to $\Reg$ for every unit multiplies every eigenvalue $\lambda_b$ by
$\mathbb E_P\chi_b(P)$, which is $0$ for $b\neq I$. The fingerprint then vanishes exactly, without calibration and
without knowing $\astar$; the attacker needs only single-qubit Paulis and random bits. If the random layer is biased,
with $\mathbb E_P\chi_{\astar}(P)=\beta$, the residual strength is $\eps\beta$, and since $\Nn$ grows as the inverse
square of the effective strength, attribution then costs about $\beta^{-2}$ times as many units. If the layer draws
the Pauli of each cell independently, with bias $\beta_i=\mathbb E\,\chi_{\astar_i}(P_i)$ on the local factor of
$\astar$ at cell $i$, then $\beta=\prod_{i\in\mathrm{supp}\,\astar}\beta_i$ decays exponentially with the weight
$|\astar|$, so the high-weight regime in which the separation is exponential is also the regime in which even a sloppy
random layer erases the fingerprint best. Appendix~B gives a
six-qubit circuit with exactly this structure for $r=3$: five tool qubits apply uniform generators of the
centralizer of $\astar$, and a sixth applies one anticommuting Pauli with probability $(1-s\eps)/2$. The support of
$\astar$ is the set of cells the tool couples coherently, and its Pauli types reflect the coupling. What the report
names is therefore a fingerprint, not a device. Tools that leave the same $(\astar,s)$ on the same cells, with
strengths that $\hat\eps$ cannot separate, are indistinguishable, and a tool built to reproduce another's fingerprint
is attributed to that tool.

Two cautions belong here. First, the uniform family, $\astar$ drawn uniformly from all $4^r-1$ non-identity Paulis
on $\Reg$, is the hard instance of Pauli channel learning \cite{chen2024tight}. It is the worst case for verifiers
without a reference. A real tool whose symmetry is local has a low-weight $\astar$, and for weight at most three we
observe no exponential gain in simulation (Section~\ref{sec:res-weight}). Second, if the attacker erases the parity ($\eps=0$),
no verifier can attribute the attack, and tier~1 still detects and localizes it. Forensics therefore concerns
attackers who do not erase their fingerprint: an attacker unaware of QMF, one whose randomization is biased, or a tool
without single-qubit control of the cells. For example, a hub insider who opens items with a supplied resealing
device has no single-qubit control of the cells: the device couples to them as built, and adding a random layer would
take a second quantum device and fresh randomness for every unit. An attacker who expects the seal to be checked by
error rates alone has no reason to add a layer that leaves those rates unchanged. As a further boundary case we consider \emph{local} tampering, which
applies a fixed Pauli $\astar$ with probability $q$ and is not scrambled.

Benign noise is independent depolarizing noise of rate $p_0$ on every seal cell during storage and transport. The
vault half carries noise $p_v$, equal to $p_0$ unless stated. The vault therefore sees the fingerprint attenuated to
$\eps_{\mathrm{eff}}=\eps(1-4p_0/3)^{|\astar|}(1-4p_v/3)^{|\astar|}$, and a verifier that reads the seal half only sees
$\eps(1-4p_0/3)^{|\astar|}$. At a fixed per-qubit noise $\eps_{\mathrm{eff}}$ shrinks exponentially with the weight,
so every noisy statement below is a finite-size statement for the simulated range $r\le10$.

\subsection{The forensics task}
\label{sec:task}
\begin{definition}[Forensics task]
\label{def:task}
The input is a record of $N$ units in arrival order, $E_{u,c}$ for $u=1,\dots,N$ and $c=1,\dots,n$, together with the
benign rate $p_0$. The output is a status code and
\begin{enumerate}
\item tier~1: the tampered region $\hat\Reg$, its error rates and the onset $\hat t_0$;
\item tier~2: the fingerprint $(\hat a,\hat s,\hat\eps)$;
\item tier~3: whether and at which unit the tool changed, and links to earlier incidents.
\end{enumerate}
Tier~2 succeeds if $(\hat a,\hat s)=(\astar,s)$. The cost of a verifier is $\Nn$, the smallest $N$ at which its tier-2
success probability, averaged over the instance prior and the randomness of data and verifier, is at least $0.9$.
Tier~3 is scored by detection rate, false-alarm rate and the error of the switch time.
\end{definition}
In most experiments $\Reg$ is given to the decoder or restricted to a stated candidate space; we call that a
\emph{conditional} recovery. The full report (Section~\ref{sec:report}) solves Definition~\ref{def:task} from raw
records. Table~\ref{tab:notation} collects the notation.

Tier~1's onset and tier~3's switch are change points in arrival order. They date the attack only if the network
delivers the items in the order the attacker touched them, as when the hub forwards them first in, first out.
Reordering in transit that interleaves the two tools' items leaves no single change point, so tier~3 detects a change
of fingerprint in arrival order; Section~\ref{sec:res-tier3} tests in simulation how much reordering it tolerates.
The full report also assumes that every unit after the onset is tampered. Clean units interleaved with tampered
ones, from reordering or from an attacker who touches only part of a consignment, bias tier~2; E15, which interleaves
clean units as a side effect of reordering, shows the effect (Sections~\ref{sec:res-tier3}
and~\ref{sec:limitations}).

\begin{table}[t]
\caption{Notation.}
\label{tab:notation}
\centering
\setlength{\tabcolsep}{3pt}
\begin{tabular}{p{1.25cm}p{6.6cm}}
\toprule
Symbol & Meaning \\
\midrule
$n$, $r$ & cells per seal unit; tampered cells, $r=|\Reg|$ \\
$u$, $N$ & seal unit (one module on one item); number of units \\
$N_{\mathrm{post}}$ & units from the onset $\hat t_0$ on (tier-2 and tier-3 statistics) \\
$E_u$ & Pauli error of unit $u$ read by the Bell measurement \\
$\astar$, $s$, $\eps$ & fingerprint, its sign and its strength, Eq.~(\ref{eq:scramble}) \\
$w$ & weight $|\astar|$ of the fingerprint \\
$p_0$, $p_v$ & benign depolarizing noise of a seal cell, of a vault qubit \\
$\eps_{\mathrm{eff}}$ & strength after benign attenuation \\
$F$ & secret Pauli frame (two classical bits per cell and unit) \\
$\Nn$ & units at which 90\% of trials identify $(\astar,s)$ \\
$T$ & time from issue to verification of a unit \\
$\ceff(r,w)$ & best coverage of weight-$w$ Paulis by one stabilizer group, Eq.~(\ref{eq:coverage}) \\
$\tau(x)$ & tier-1 threshold $x+5\sqrt{\max(x,1/N)/N}+5/N$ \\
$W$ & tier-3 window, in units \\
\bottomrule
\end{tabular}
\end{table}

\subsection{Verifier classes}
The comparison is fixed by what a verifier can prepare at issue and what it can measure together after the attack
(Table~\ref{tab:classes}). A per-unit verifier has \emph{no untouched reference}: every qubit of every probe it
prepares lies on a cell of $\Reg$ (Definition~\ref{def:perunit} in Section~\ref{sec:theory}).

\begin{table*}[t]
\caption{Verifier classes. Every result in this paper names its class. ``Within reach'' means on a cell the attacker
can act on.}
\label{tab:classes}
\centering
\setlength{\tabcolsep}{4pt}
\begin{tabular}{L{2.6cm}L{5.6cm}L{3.9cm}L{4.2cm}}
\toprule
Class & Allowed resources & Deployment example & Implemented strategies \\
\midrule
Per-unit & Entanglement inside a unit, classical records, adaptive control; no quantum coupling between units; every probe
qubit within reach & Items verified and released on arrival, in an order the verifier does not choose &
Stabilizer probes (uniform Lagrangian, product, mixed, GHZ, adaptive GHZ); product probe of copies \\
Self-referencing & Entanglement and joint measurement across two units; single-unit probes after learning & Pairs of
items fixed at issue, held until both arrive, measured jointly & P (cell-wise Bell pairs across two units); P + sign \\
In-seal reference & Per-unit resources plus untouched seal cells as references & Spare cells inside one module & In-seal
Bell pairs; in-seal oracle told $\Reg$ \\
Protected reference & A quantum memory (vault) kept coherent for the seal lifetime & The verifier's vault & Bell
measurement against the vault \\
\bottomrule
\end{tabular}
\end{table*}

\emph{Why per-unit verifiers are the main comparison.} A seal cell is itself a quantum state that stays coherent from
issue to verification, so the obstacle to coupling two units is not the lack of a memory. It is logistics and
timing. A pair of units can be measured jointly only if the pairing was fixed at issue and both items are at the
inspection site at the same time. In the scenario of Section~\ref{sec:scenario} items arrive in an uncontrolled order
and are released on arrival, so the first item of a pair is gone before its partner arrives, and no pair can be
measured jointly. That makes the per-unit class the vault-free alternative. If the operator can instead hold an item
until its partner arrives, the self-referencing verifier is available, and we report it next to every per-unit
result.

\emph{Adaptive preparation versus adaptive measurement.} Some implemented strategies choose how later units are
\emph{prepared} from the results of earlier ones: the adaptive stabilizer probes, the adaptive in-seal pairing and the
sign phase of P + sign. When a whole consignment is issued before any item is read, as in our scenario, these
strategies need a second consignment. The order, bases and stopping rule of the measurements can always adapt.

\subsection{Attackers}
Table~\ref{tab:attackers} lists the attackers. All are non-adaptive in the sense that the same channel acts on every
unit, and none knows the frame. The \emph{oblivious} attacker applies (\ref{eq:scramble}) to $\Reg$ and nothing else.
\emph{Extend} also fully depolarizes every cell outside $\Reg$; it needs no secret and destroys every untouched
reference. \emph{Spray} splits the cells outside $\Reg$ into $r$ groups and copies the error of tampered cell $i$
onto every cell of group $i$; it needs no secret. \emph{Mirror} copies each tampered cell's error onto that cell's
reference partner, so it must know the pairing.

\begin{table}[t]
\caption{Attackers. All act with the same channel on every unit and do not know the Pauli frame.}
\label{tab:attackers}
\centering
\setlength{\tabcolsep}{3pt}
\begin{tabular}{L{1.5cm}L{2.75cm}L{1.2cm}L{2.0cm}}
\toprule
Attacker & Acts on & Secret needed & Experiments \\
\midrule
Oblivious & $\Reg$ & none & E1--E9 \\
Extend & $\Reg$; depolarizes all other cells & none & E9; bounds \\
Spray & $\Reg$; copies errors to groups & none & E11, E12 \\
Mirror & $\Reg$ and partners & pairing & E9 \\
\bottomrule
\end{tabular}
\end{table}

\emph{An attacker who erases the fingerprint.} Forensics presumes that one tool acts on many units and leaves its
fingerprint. An attacker who knows QMF can remove the fingerprint at little cost: a uniformly random Pauli on $\Reg$
after the tool, drawn afresh for each unit, averages every non-identity eigenvalue to zero (Section~\ref{sec:tamper}).
Mixing tools is a weaker form of the same averaging. An attacker who picks one of $K$ tools at random for each unit
produces a single channel with $K$ hidden parities of strength $\eps/K$ each, and the random Pauli layer is the limit
of a mixture over all $4^r-1$ parities. Our several-parity experiment E7 samples exactly the $K$-tool mixture for
$K\le3$ and $r\le8$, with $\eps$ and $K$ unknown to the decoder. The vault identifies all $K$ parities in every tested
setting, at a cost that grows roughly as $K^2$, and the per-unit strategy stays \qnEsevenRatio{} times more expensive at
$r=8$ (Supplement~S3). An erased fingerprint leaves tiers~2 and~3 nothing to read for any verifier, vault included,
while tier~1 still detects and localizes the tampering because the cells' error rates stay high. This is a
fundamental limit of the method: QMF attributes attackers who do not erase their fingerprint, and erasing it is cheap.
Leakage of the frame and changes of strength alone are also outside the model.

\subsection{Implementation levels}
We separate three levels of evidence. At the first, Qiskit circuits of the issue, tamper, probe and verification
blocks reproduce the sampler's distributions exactly \cite{qiskit2024} (\qnCircuitChecks{} checks in all, to within
$\qnCircuitMaxDev$; Supplement~S5). At the second, the pair $(\astar,s)$ is decoded with $\Reg$ given or within a
stated candidate space; most experiments are at this level. At the third, the full report goes from raw vault records
to location, onset, fingerprint and switch (Section~\ref{sec:report}). A success at the second level is never
reported as a success of the full report.

\section{The QMF Report}
\label{sec:report}

The report (Algorithm~\ref{alg:qmf}) solves the forensics task of Definition~\ref{def:task} from the vault's
Bell-measurement records, a units-by-cells array of Paulis in arrival order, and the benign rate $p_0$. It returns a
status code and the three tiers. Both tier-1 tests use the same threshold function
$\tau(x)=x+5\sqrt{\max(x,1/N)/N}+5/N$, five binomial standard deviations above a rate $x$ plus a small-sample guard.

\begin{algorithm}[t]
\caption{QMF report (full pipeline)}
\label{alg:qmf}
\begin{algorithmic}[1]
\REQUIRE records $E_{u,c}$ ($u=1..N$ units, $c=1..n$ cells), benign rate $p_0$, window $W$, optional registry
\STATE $\rho_c\gets$ fraction of units with $E_{u,c}\neq I$; \ $\tilde\rho\gets\mathrm{median}_c\,\rho_c$;
\ $b_0\gets 1-(1-p_0)^2-p_0^2/3$
\IF{$\tilde\rho>\tau(b_0)$}
  \RETURN \texttt{global\_anomaly} (more than half of the cells look tampered; not localized)
\ENDIF
\STATE $\hat\Reg\gets\{c:\rho_c>\tau(\tilde\rho)\}$ \COMMENT{tier 1: where, how much}
\IF{$\hat\Reg=\emptyset$} \RETURN \texttt{clean} \ENDIF
\STATE onset $\hat t_0\gets$ Bernoulli change point of ``any error on $\hat\Reg$'' \COMMENT{tier 1: from when}
\IF{two cells of $\hat\Reg$ agree on $\ge 0.8$ of the units where either shows an error}
  \RETURN \texttt{copied\_errors} (fingerprint ambiguous; not decoded)
\ENDIF
\IF{$|\hat\Reg|>12$} \RETURN \texttt{not\_decoded} \ENDIF
\STATE $\hat\lambda_b\gets$ empirical parity of every $b$ on $\hat\Reg$ over the $N_{\mathrm{post}}$ units from $\hat t_0$
on (Walsh--Hadamard, $O(r4^r)$)
\STATE $(\hat a,\hat s,\hat\eps)\gets\arg\max_{b\neq I}|\hat\lambda_b|$; significant if
$|\hat\lambda_{\hat a}|>\sqrt{2\ln(2\cdot4^r/0.01)/N_{\mathrm{post}}}$; if not, the fingerprint is returned flagged as
not significant and is not used (no attribution from the blind search) \COMMENT{tier 2}
\STATE fingerprints of the first and last $W$ units (same maximization over $b\neq I$ and same test, $W$ in place of
$N_{\mathrm{post}}$); if both are
significant and differ in $a$, locate the switch by a likelihood-ratio change point whose gain exceeds
$2\ln N_{\mathrm{post}}$ \COMMENT{tier 3}
\STATE link to registry entries with the same cells by a known-signature test ($z>4$ on each registry fingerprint; it
does not use the decoded fingerprint and runs whether or not tier~2 is significant)
\RETURN \texttt{localized}, tiers 1--3, and \texttt{tier3} $\in$ \{\texttt{switch}, \texttt{no\_switch},
\texttt{unsupported\_same\_a}, \texttt{not\_tested}\}
\end{algorithmic}
\end{algorithm}

\subsection{Tier 1: where, how much, from when}
The report first compares the median cell error rate with the benign rate $b_0$ of a Bell outcome whose two halves
are each depolarized at rate $p_0$. A median above $\tau(b_0)$ means that more than half of the cells look tampered.
The region cannot then be localized from rates, and the report returns \texttt{global\_anomaly}; extend and spray both
trigger it. Otherwise the tampered cells are those whose rate exceeds $\tau(\tilde\rho)$, and a Bernoulli change point
on ``any error in $\hat\Reg$'' gives the onset. Error rates can be estimated without entanglement from few units
\cite{chen2022quantum,flammia2021pauli}, so tier~1 offers no separation between verifier classes. We did not simulate
tier~1 for per-unit verifiers.

\subsection{Tier 2: fingerprint}
On the units after the onset, a fast Walsh--Hadamard transform of the empirical distribution of $E$ restricted to
$\hat\Reg$ gives every parity $\hat\lambda_b$ at once, at a cost of $O(r4^r)$. The report takes the profile
maximum-likelihood fingerprint $\hat a=\arg\max_{b\neq I}|\hat\lambda_b|$ (the identity's empirical parity is always
$+1$), its sign and its attenuation-corrected strength, and declares it significant only if $|\hat\lambda_{\hat a}|$
clears a Bonferroni threshold at 1\% over the $4^r$ candidates. If the test fails, the status stays
\texttt{localized} with the tier-1 results, the fingerprint is returned flagged as not significant, and no fingerprint
is attributed from the blind search; tier~3 runs only if both window fingerprints pass their own test, and registry
links, which test known fingerprints, still run. The experiments use exact Bayesian decoding over $(a,s)$ with $\eps$ and $p_0$ known
(Section~\ref{sec:setup}). Decoding is exponential in $r$ in classical time, and the report stops at $r=12$
(\texttt{not\_decoded}). For larger regions the problem is noisy sparse-parity learning \cite{blum2003noise}, which
the vault solves in samples but not necessarily in time.

Before decoding, the report checks for copies. If two cells of $\hat\Reg$ agree on at least 0.8 of the units where
either shows an error, the cells carry copies of each other's errors, as under mirror. Which cell is the original,
and hence $\astar$, is then ambiguous, and the report returns \texttt{copied\_errors} without decoding. Counting only
units with an error keeps two rarely hit cells from looking like copies. Supplement~S4 gives the reasons for each
threshold.

\subsection{Tier 3: when the tool changed, and links}
The report computes the fingerprints of the first and the last $W$ units, with the same maximization over $b\neq I$
and the same test with $W$ in place of $N_{\mathrm{post}}$. If both are significant and their Paulis differ, a likelihood-ratio change point locates the switch. The supported scope is one switch between tools with
different $\astar$. A change of sign with the same $\astar$ is reported as \texttt{unsupported\_same\_a}. A change of
strength alone, or several changes, is not detected and was not evaluated. Links to a registry of earlier incidents
use a known-signature test on the same cells; a link means the same fingerprint on the same cells, not the same
device. That test needs no memory; it is implemented and unit-tested but not
evaluated on populations of incidents.

\subsection{Supported scope}
The full report supports an oblivious tool on fewer than half of the cells, with at most one switch between tools with
different fingerprints, in a record in which every unit after the onset is tampered. Every report carries a status and a \texttt{tier3} field. The vault's fingerprints under
extend, spray and mirror in Section~\ref{sec:results} come from separate, conditional decoders and are not successes
of this report.

\subsection{A worked incident}
\label{sec:case}

\begin{figure*}[t]
\centering
\includegraphics[width=\textwidth]{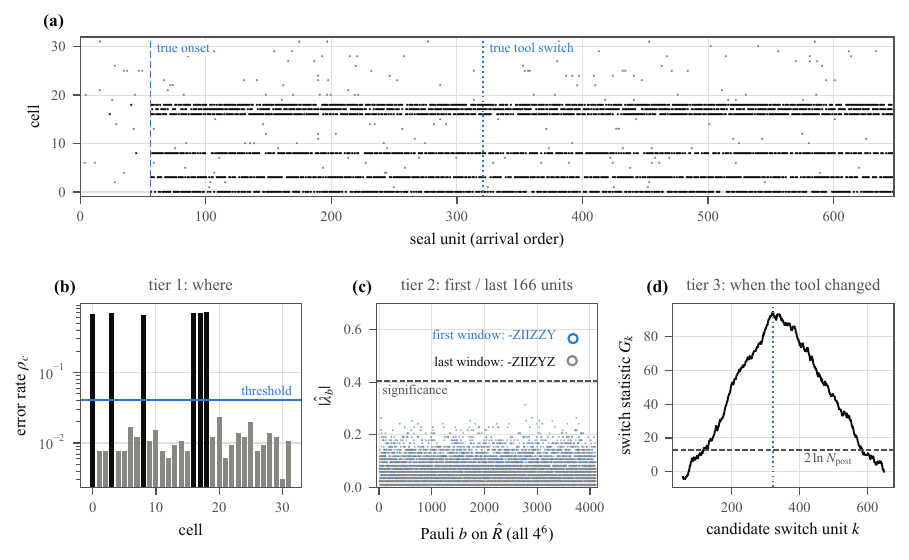}
\caption{One incident from raw records to the report: an E6 vault trial ($r=6$, $\eps=0.6$, $p_0=0.005$, 32 cells,
\qnCaseUnits{} units), chosen before it was drawn (trial 0, a switch trial).
(a)~Non-identity Bell outcomes by unit and cell; black marks the tampered cells. Lines: true onset and true tool switch.
(b)~Tier~1: cell error rates and the threshold $\tau(\tilde\rho)$.
(c)~Tier~2: $|\hat\lambda_b|$ for the $4^6$ Paulis, with the identity, whose parity is always 1, set to 0 and
excluded from the maximization, in the first (blue) and the last (gray) $\qnCaseWindow$ units after the onset; circles mark the two decoded fingerprints, the dashed line is the significance threshold for a window of
\qnCaseWindow{} units.
(d)~Tier~3: the switch statistic $G_k$, its threshold $2\ln N_{\mathrm{post}}$, and the true switch (dotted).}
\label{fig:case}
\end{figure*}

Fig.~\ref{fig:case} follows one incident through the report. The record has \qnCaseUnits{} units: a benign stretch,
then tool~1 and then tool~2 on the same six cells. In panel~(a) the tampered cells stand out once the attack starts,
and nothing in the rates marks the change of tool, because both tools scramble every cell completely. Tier~1 finds the
six cells and the onset at unit \qnCaseOnset{} exactly (panel~b). Tier~2 then reads the full error samples of those
cells. The first and the last \qnCaseWindow{} units each show one parity far above the rest, \qnCaseToolOne{} and
\qnCaseToolTwo{}, the two tools' fingerprints (panel~c). Because the two fingerprints differ, tier~3 runs the change
point, which peaks at unit \qnCaseSwitchFound{} against a true switch at \qnCaseSwitchTrue{} (panel~d). The record is a
vault record. A per-unit verifier sees the same rates, but in place of panel~(c) it holds syndromes of $r$ parities
per unit, and at this budget it cannot single out either fingerprint (Section~\ref{sec:res-tier3}).

\section{Separations and Bounds}
\label{sec:theory}

This section states what is proved and under which conditions. A ratio is called \emph{proved} only if it divides a
lower bound by an upper bound. A lower bound divided by a simulated cost locates the bound but proves nothing; we report
such ratios separately. Full proofs are in Appendix~A.

\subsection{Per-unit verifiers: reduction to a known lower bound}
\begin{definition}[Per-unit verifier]
\label{def:perunit}
A per-unit verifier prepares, for each unit, a state of qubits that all lie on cells of $\Reg$ (it holds \emph{no
untouched reference}), lets the tamper channel act once, and measures. It may use entanglement inside the unit,
classical memory, and adaptive choices of later preparations and measurements, but no quantum coupling between units.
\end{definition}
Per unit, this is one use of the tamper channel by an entanglement-free protocol in the sense of
Ref.~\cite{chen2024tight}: any states and measurements on the $r$ cells, entangled among them or not, with classical
memory and adaptivity but no ancilla.

\begin{proposition}[Per-unit lower bound]
\label{prop:chen}
Let the attacker apply (\ref{eq:scramble}) on $\Reg$ and nothing else, with $\astar$ uniform over the $4^r-1$
non-identity Paulis on $\Reg$, $s$ uniform, $\eps\le 1/3$ and any benign noise $p_0\ge0$. A per-unit verifier whose
success probability in identifying $(\astar,s)$, averaged over this prior, is at least $0.9$ needs
$N=\Omega(2^r/\eps^2)$ units.
\end{proposition}

\emph{Correspondence with the cited theorem.} Theorem~2 of Ref.~\cite{chen2024tight} states that an entanglement-free
protocol that estimates any requested Pauli eigenvalue of an $n$-qubit Pauli channel to precision
$\eps_{\mathrm C}\le1/6$ with probability at least $2/3$ needs $\Omega(2^n/\eps_{\mathrm C}^2)$ rounds of measurement,
with adaptivity and classical memory allowed. Its proof (Supplemental Material, Sec.~S3,
Eqs.~(S13)--(S15)) uses the family $\Lambda_{a,\pm}=|\sigma_0\rangle\!\rangle\langle\!\langle\sigma_0|
\pm\eps_0|\sigma_a\rangle\!\rangle\langle\!\langle\sigma_a|$, $P_a\neq I$, against the completely depolarizing
channel $\Lambda_0$, with $\eps_0\le1/3$ and $\eps_{\mathrm C}=\eps_0/2$. A referee draws $a\neq I$ and $s=\pm1$
uniformly and sends $N$ uses of either $\Lambda_0$ or $\Lambda_{a,s}$ with equal probability; the bound follows from
the two-point inequality on the total-variation distance averaged over $a$ (S15). The Pauli eigenvalue of
$\Lambda_{a,s}$ at $a$ is $s\eps_0$, so our $(\astar,s,\eps)$ is their $(a,s,\eps_0)$ with $n=r$. The condition is
therefore $\eps\le1/3$ in our parametrization, not the precision bound $\eps_{\mathrm C}\le1/6$. Of our two strengths,
$\eps=0.15$ lies inside this range and $\eps=0.6$ does not.

\begin{proof}[Sketch]
From an identifier $\mathcal A$ we build a player for the referee's game: run $\mathcal A$ on $N$ units, then spend
$m=\lceil 8\ln 20/\eps^2\rceil$ more units measuring the eigenvalue of $\hat a$, and declare $\Lambda_{a,\pm}$ if the
average is closer to $\hat s\eps$ than to~$0$. The player uses only per-unit resources and wins with probability at
least $0.9$ averaged over the prior, so the averaged distance in (S15) is at least $1/3$, and the rest of the proof of
Theorem~2 gives $N+m=\Omega(2^r/\eps^2)$. Benign noise cannot help: the verifier could apply the same depolarizing
noise itself before and after the attack, so a protocol for noise $p_0$ yields one for $p_0=0$ with the same $N$.
Appendix~A gives the details.
\end{proof}

The proposition needs three conditions. The verifier must be per-unit, with no untouched reference. The channel must
be the tool's own: an attacker that copies errors onto other cells changes the channel into an easy one
(Section~\ref{sec:res-attack}). And the fingerprint must be one parity with $\eps\le1/3$. For $\eps=0.6$, and for
several parities with unknown strength, our evidence is simulated (Section~\ref{sec:results}).

The constant is explicit. Inserting the distance bound of Ref.~\cite{chen2024tight} (Supplemental Material,
Eqs.~(S67) and (S69)) into the reduction gives, for every per-unit verifier,
\begin{equation}
N\ \ge\ \frac{0.8\,(4^r-1)}{2^r\eps^2\bigl(1+2\sqrt{f(\eps)}\bigr)}-\Bigl\lceil\frac{8\ln20}{\eps^2}\Bigr\rceil ,
\label{eq:chen-explicit}
\end{equation}
with $f$ given in Appendix~A ($f<44$ for $\eps\le1/3$). At $r=10$ and $\eps=0.15$ this is \qnChenLbRten{} units. That
exceeds the vault's simulated $\Nn$ of \qnVaultNninetyRtenLow, which locates the bound but proves nothing, and it is
below the vault's information-theoretic upper bound of \qnChenVaultUbRten{} (Proposition~\ref{prop:ub}). Within our
simulated range the reduction therefore proves the $2^r$ scaling but not a separation against every per-unit
verifier; at $\eps=0.15$ and $p_0=0.005$ it does so from $r=\qnChenRstar$. The proved ratios at $r\le10$
(Section~\ref{sec:proved}) rest on the Fano bound below, which covers non-adaptive stabilizer probes only.

\subsection{Non-adaptive stabilizer probes: a Fano bound}
A per-unit stabilizer probe prepares a stabilizer state with group $S_u$ and reads its syndromes. This reveals the
parity of every element of $S_u$ and nothing about Paulis outside it. For a family $W_w$ of the $M=\binom{r}{w}3^w$
Paulis of weight $w$, define the best coverage
\begin{equation}
\ceff(r,w)=\max_{L}\frac{|L\cap W_w|}{M},
\label{eq:coverage}
\end{equation}
over all Lagrangian (maximal isotropic) subgroups $L$ of the $r$-qubit Pauli group.

\begin{proposition}[Fano bound]
\label{prop:fano}
Let $(\astar,s)$ be uniform over $W_w\times\{\pm1\}$, and let $\eps'=\eps(1-4p_0/3)^{w}$ be the attenuated strength.
Every non-adaptive per-unit stabilizer probe whose success probability, averaged over this prior, is at least $0.9$
needs
\begin{equation}
N\ \ge\ \frac{0.9\log_2(2M)-1}{\ceff(r,w)\,d(\eps')},\qquad d(x)=1-h_2\!\left(\tfrac{1+x}{2}\right),
\label{eq:fano}
\end{equation}
units, where $h_2$ is the binary entropy. The bound holds also when $\Reg$ is unknown.
\end{proposition}
The success probability here, as in Proposition~\ref{prop:chen}, is the average over the instance prior; that is
what $\Nn$ measures, since each trial draws its instance from the prior.

For full weight, the number of weight-$r$ elements of a stabilizer group equals the $r$-body sector length of its
state. That sector length is at most $2^{r-1}+[r\ \text{even}]$ for every $r$, with equality for GHZ
\cite{tran2016correlations,eltschka2020maximum}, so $\ceff(r,r)=(2^{r-1}+[r\ \text{even}])/3^r$. We confirmed this
independently by enumerating all Lagrangians ($r\le6$) and all labeled graph states ($r\le8$), and by an integer
program for $r=9$. For other weights the integer program gives an upper bound on $\ceff$, which keeps
(\ref{eq:fano}) valid and conservative. The stored $r=10$, $w=r$ bound uses the integer-program value 540 rather than
the theorem's 513, so it is $513/540$ of the bound the theorem allows; the theorem's value gives a 5.26\% larger
bound (Table~\ref{tab:proved}). For the uniform family a single group covers $(2^r-1)/(4^r-1)$ of the non-identity
Paulis, and the same argument with $M=4^r-1$ and that coverage gives the uniform bound of
Table~\ref{tab:proved}; there we use the unattenuated $\eps$ in $d$, which only weakens it.

\subsection{Information-theoretic upper bounds}
For outcome laws of the form $P_{a,s}(E)=4^{-n}(1+s\eps_a\chi_a(E))$, with $\chi_a(E)=(-1)^{\langle a,E\rangle}$, any
two hypotheses $(a,s)\neq(b,s')$ have Bhattacharyya coefficient
\begin{equation}
\mathrm{BC}=
\begin{cases}
g(\eps_a)\,g(\eps_b), & a\neq b,\\
\sqrt{1-\eps_a^2}, & a=b,\ s\neq s',
\end{cases}
\label{eq:bc}
\end{equation}
where $g(x)=\tfrac12(\sqrt{1+x}+\sqrt{1-x})\le 1-x^2/8$ \cite{bhattacharyya1943}. A union bound then makes maximum
likelihood err with probability at most $|H|(1-\eps_{\min}^2/8)^N$ on every instance, where $|H|$ is the number of
hypotheses.

\begin{proposition}[Upper bounds]
\label{prop:ub}
Error at most $\delta$ on every instance is reached with the following numbers of units.
\begin{enumerate}
\item Vault, $\Reg$ given:
$N_{\mathrm V}=\lceil 8\ln(2(4^r-1)/\delta)/\eps_{\min}^2\rceil$, where $\eps_{\min}=\eps(1-4p_0/3)^{2r}$.
\item Vault, $\Reg$ unknown among $n$ cells (extend): the same with $4^r-1$ replaced by
$M_n=\sum_{k=1}^{r}\binom{n}{k}3^k$; here $\ln M_n\le r\ln(3en/r)+\ln r$.
\item P + sign, $\Reg$ unknown among $n$ cells (extend): $2\lceil 8\ln(M_n/\delta_a)/\eps_P^2\rceil$ units of pairs with
$\eps_P=\eps^2(1-4p_0/3)^{2r}$, plus $\lceil 2\ln(1/\delta_s)/\eps_s^2\rceil$ sign units with
$\eps_s=\eps(1-4p_0/3)^{r}$, for $\delta=\delta_a+\delta_s$.
\end{enumerate}
\end{proposition}
These are bounds on samples, not on classical decoding time. Each carries a noise attenuation: $(1-4p_0/3)^{-4r}$ for
the vault and for the pair part of P + sign, and $(1-4p_0/3)^{-2r}$ for its sign part. Each bound is therefore
polynomial in $r$ at zero noise or at a fixed effective bias, but not at a fixed positive per-qubit noise. We evaluate
them at $p_0=0.005$ and $r\le10$ with $\delta=0.1$ ($\delta_a=0.09$, $\delta_s=0.01$).

\subsection{Proved ratios}
\label{sec:proved}
Table~\ref{tab:proved} divides the Fano bound (\ref{eq:fano}) by the upper bounds of Proposition~\ref{prop:ub}. Where
the result exceeds one, the protected reference (or P + sign) provably needs fewer units than every non-adaptive
per-unit stabilizer probe. Under extend the Fano bound applies with $\Reg$ unknown, because the informative stabilizer
elements form an isotropic subgroup on $\Reg$ (Lemma~\ref{lem:extend}, Appendix~A).

\begin{table}[t]
\caption{Proved ratios at $r=10$, $p_0=0.005$: Fano lower bound (non-adaptive per-unit stabilizer probes) over an
information-theoretic upper bound. Ranges are over $\eps\in\{0.15,0.6\}$ and are rounded down.}
\label{tab:proved}
\centering
\setlength{\tabcolsep}{3pt}
\begin{tabular}{p{3.6cm}p{2.0cm}r}
\toprule
Family and attacker & Upper bound & Ratio $\ge$ \\
\midrule
Uniform, oblivious & vault, $\Reg$ given & \qnProvedUniform \\
$w=r$, oblivious & vault, $\Reg$ given & \qnProvedWr \\
\quad with the theorem's $\ceff$ & & \qnProvedWrThm \\
$w=\lceil r/2\rceil$, oblivious & vault, $\Reg$ given & \qnProvedWhalf \\
Uniform, extend & vault, $\Reg$ unknown & \qnProvedExtendPerUnit \\
Uniform, extend, $\eps=0.6$ & P + sign, $\Reg$ unknown & \qnProvedPsignRten{} ($r=8$: \qnProvedPsignReight) \\
\bottomrule
\end{tabular}
\end{table}

The same lower bounds divided by the \emph{simulated} vault $\Nn$ are not proofs. They are \qnLbOverMeasUniform{} for
the uniform family (fine checkpoints), \qnLbOverMeasWr{} for $w=r$ and \qnLbOverMeasWhalf{} for $w=\lceil r/2\rceil$.
They are larger because the simulated vault is cheaper than its information-theoretic bound. At $\eps=0.15$ the
P + sign bound is not small enough to prove an advantage over per-unit probes under extend. For $w\le2$ the Fano bound
is below the vault's upper bound, so no separation is proved there.

\section{Evaluation Setup}
\label{sec:setup}

\subsection{Instances and laboratories}
Every experiment draws its instances from one seed space, so different verifiers face the same fingerprints. The
fingerprint $\astar$ is uniform over the non-identity Paulis on $\Reg$, or uniform over a weight family, and the sign
is uniform. We use $\eps\in\{0.15,0.6\}$ and $p_0\in\{0,0.005\}$ with $r=4$ to $10$ tampered cells (E14 raises the
seal noise to $p_0=0.1$ with the vault's noise held at $0.005$; Section~\ref{sec:engineering}); $\eps=0.15$ lies
inside the range of Proposition~\ref{prop:chen} ($\eps\le1/3$) and $\eps=0.6$ does not. Every decoder knows $\eps$
and $p_0$ and computes the exact posterior over $(a,s)$; success means $(\hat a,\hat s)=(\astar,s)$.

The laboratories follow the verifier classes. The vault lab draws Bell outcomes from their exact law, with the
attenuation of both halves. The per-unit labs prepare stabilizer probes from uniformly random Lagrangians, product
states, a mixture of the two, GHZ-type states, and an explore/exploit adaptive GHZ heuristic; each unit's syndrome
updates the exact likelihood. The self-referencing labs are P, cell-wise Bell pairs across two units, which reveal
$E_uE_{u'}$ with bias $\eps^2$ and no sign, and P + sign, which learns $\hat a$ with P and then spends
$m=\lceil 2\ln100/(\eps(1-4p_0/3)^r)^2\rceil$ single units on the eigenvalue of $\hat a$ to learn the sign. A pair
counts as two units, and the sign phase is included in the cost. The full-seal labs simulate a 32-cell seal with
in-seal Bell references, the vault with its own tier~1 and copy merging, a reference-free product probe that reads
copies in complementary bases, and the attackers of Table~\ref{tab:attackers}.

\subsection{Sample cost}
All empirical results come from simulation; there is no hardware run. The cost of a verifier is $\Nn$
(Definition~\ref{def:task}). Each trial is decoded at 40 geometric checkpoints, and $\Nn$ is read on a common grid of
40 geometric points. Success ``at the horizon'' means success at a trial's last checkpoint, its final simulation
horizon, which lies beyond $\Nn$; it is not the success fraction at $\Nn$. We give 90\% bootstrap intervals over trials and 95\%
Clopper--Pearson intervals for rates. Trial counts per setting: in E1 the vault has 200 and the per-unit baseline 50
(20 at $r=9,10$); in the weight families (E2, E5b) every laboratory has 50 (20 at $r=10$); P, P + sign and local
tampering have 20; the vault-noise grid (E3) 50; several parities (E7) 100 (vault) and 30--50 (per-unit); tier~3 300
switch and 300 no-switch trials (vault) and 100 and 100 (per-unit); the full-seal attack experiments (E9, E11, E12) 50
to 200; and E14 200 (vault) and 50 (per-unit). A setting whose success never reaches 0.9
is censored and reported as a lower bound. When the trials of a group have different horizons the grid reads $\Nn$
late, so for the uniform family we replayed the trials with the same draws at fine checkpoints; headline numbers give
the fine value first and the grid value in parentheses, and other grid values from such groups carry ``$\pm1$ grid
step''. E14 avoids the effect by design: every trial of a group shares the group's fine checkpoints. Appendix~C gives
the procedure.

\subsection{Pre-registration and reproducibility}
The results come from one line of runs on the final code, which supersedes an earlier line; the earlier line is not
used for any claim. The protocol, the predictions and the decision rule of every experiment were
written into a laboratory notebook, and analyses added after seeing results are marked post hoc there and in the
experiment register (Supplement~S1). The register lists each experiment's question, rule, outcome, trial count and
seed space, and sets the time in its notebook heading next to the start time that the run script logged. For E1f
and E11--E15 the heading carries a clock time that precedes the logged start. For the other experiments from E3 on it
carries only the date, which matches the run date but cannot order the two within the day. For E1 and E2 the logged
runs start the day before the date in the notebook heading, so we do not claim
documented pre-registration for them; their rules and outcomes are reported as written. Four items are post hoc: the
GHZ-type per-unit baseline (its prediction and rule were written before it ran, after the first weight results), the
analysis fix of the vault-noise crossover, the rerun of P and of local tampering, and the reordering test E15 (its
protocol and rules were written before it ran, after the other results).
Failed predictions are reported as failed. The notebook is available from the authors on request; its times are
self-recorded, and Supplement~S1 sets each against the start time that the run logged. The public repository and its
Zenodo deposit (doi:~10.5281/zenodo.24148790, published 2026-11-12) time-stamp the code and the results.

All raw rows are stored with their seeds, and a check script regenerates sample rows from the seeds. The result figures
and tables and the headline numbers of this paper are rebuilt from the raw rows: every summary they need is recomputed
in memory by the analysis code, and no stored summary is read. The repository carries the SHA-256 of every file and
passes the same checks in two software environments, one of them with Qiskit for the circuit checks \cite{qiskit2024}. Supplement~S7 gives the
commands, environments and running times.

\section{Results}
\label{sec:results}

\subsection{Uniform fingerprint: three strategies}
\label{sec:res-uniform}

\begin{figure*}[t]
\centering
\includegraphics[width=\textwidth]{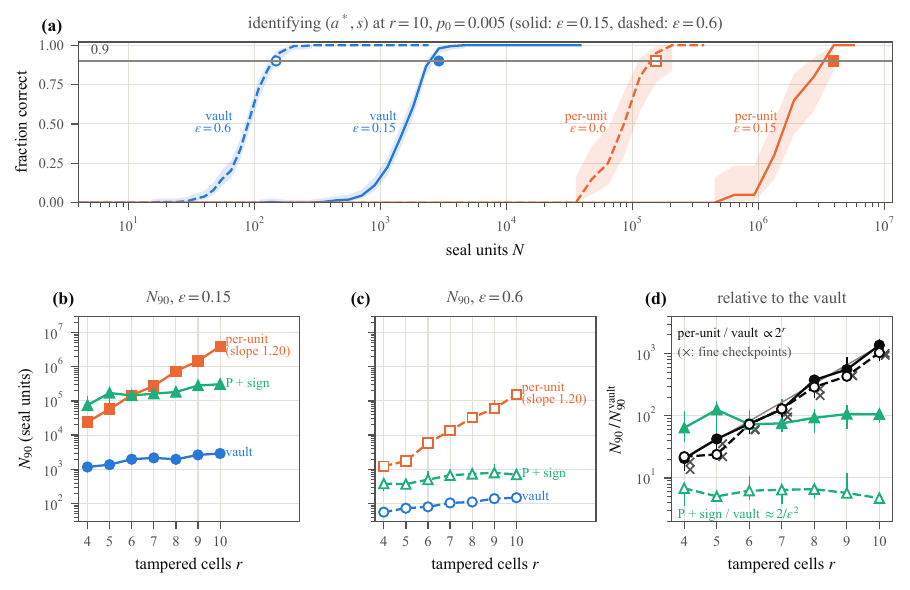}
\caption{Uniform hidden-parity fingerprint at $p_0=0.005$: the vault against the per-unit strategy and the
implemented self-referencing strategy.
(a)~Fraction of trials that identify $(\astar,s)$ against seal units at $r=10$. Bands are 95\% Wilson intervals over
the transition (200 vault and 20 per-unit trials); markers are $\Nn$.
(b), (c)~$\Nn$ with 90\% bootstrap intervals at $\eps=0.15$ and $\eps=0.6$ for the vault, the per-unit strategy
(uniform Lagrangian probes; the label gives the $\log_2$ slope in $r$) and P + sign (20 trials; a pair counts as two
units, and the sign phase is included). At $\eps=0.15$ the per-unit strategy is cheaper for $r\le\qnCrossR$ and
P + sign above; this is a crossover between two implemented strategies, not between classes.
(d)~$\Nn$ over the vault's $\Nn$ (filled: $\eps=0.15$, open: $\eps=0.6$). Black: per-unit, $\propto2^r$ (gray line);
crosses: the same on fine checkpoints. Aqua: P + sign, flat near $2/\eps^2$; a simulated ratio for this strategy, not
a bound over the self-referencing class.}
\label{fig:uniform}
\end{figure*}

The vault's cost grows nearly linearly in $r$, while the per-unit strategy's grows as $2^r$
(Fig.~\ref{fig:uniform}). At $r=10$ and $p_0=0.005$ the vault needs \qnVaultNninetyRtenLow{} units at $\eps=0.15$ and
\qnVaultNninetyRtenHigh{} at $\eps=0.6$, and the per-unit strategy needs \qnEoneRatioFine{} times as many on fine
checkpoints (grid: \qnEoneRatioGrid), over both strengths and both noise levels. This is what
Proposition~\ref{prop:chen} predicts for $\eps\le1/3$, and against non-adaptive stabilizer probes the ratio is proved
to be at least \qnProvedUniform{} (Table~\ref{tab:proved}). The separation also survives a harder decoding problem:
with up to three hidden parities of unknown strength, which is also the channel of an attacker who switches among up to
three tools at random, the ratio at $r=8$ is \qnEsevenRatio{} ($\pm1$ grid step; Supplement~S3). In the scenario of
Section~\ref{sec:scenario} these numbers mean that an attacker can open about a thousand times more items with one
tool before a per-unit verifier attributes the tool than before the vault does.

The implemented self-referencing strategy closes most of that gap. P + sign learns $(\astar,s)$ in every setting
(\qnPsignSuccess{} trials correct at each trial's final simulation horizon, which lies beyond $\Nn$; this is not the
success fraction at $\Nn$), and its sign phase costs only \qnPsignSignCost{} times the $\Nn$ of
P alone. Its ratio to the vault is flat in $r$ at both noise levels: \qnPsignOverVaultFineLow{} at $\eps=0.15$ and
\qnPsignOverVaultFineHigh{} at $\eps=0.6$ against the fine vault $\Nn$ (grid: \qnPsignOverVaultGridLow{} and
\qnPsignOverVaultGridHigh), close to $2/\eps^2$, which is 89 and 5.6. The reason is that a pair reveals the product of
two error samples, whose parity has bias $\eps^2$ instead of $\eps$. Where items can be held until their partners
arrive, the vault's gain over this strategy is therefore a simulated polynomial factor. We claim neither the optimal
cost of the self-referencing class nor a lower bound on the vault's advantage over it.

\subsection{Fingerprint weight and stabilizer coverage}
\label{sec:res-weight}

\begin{figure*}[t]
\centering
\includegraphics[width=0.78\textwidth]{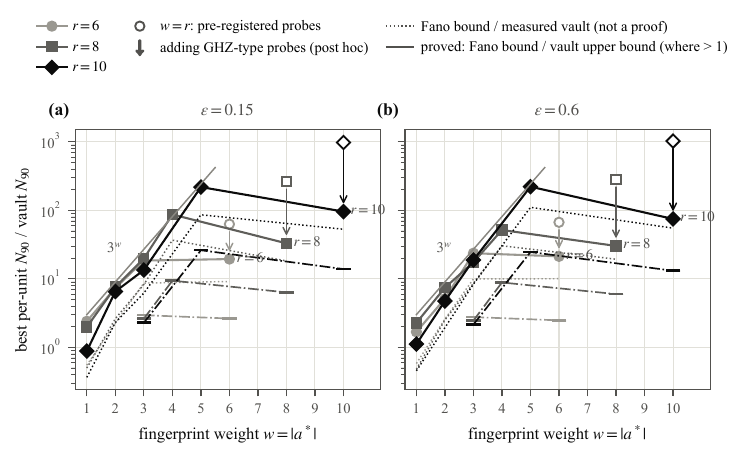}
\caption{Fingerprint weight at $p_0=0.005$. The $y$ axis is the best per-unit $\Nn$ over the vault's $\Nn$, among
product, uniform-Lagrangian and mixed probes. At $w=r$, open markers use the pre-registered probes only; filled markers
add GHZ-type probes (post hoc) and the explore/exploit adaptive GHZ probe. Dotted: the Fano bound over the
\emph{simulated} vault $\Nn$, which locates the bound but proves nothing. Dash-dot with bars: the proved ratio, Fano
bound over the vault's information-theoretic upper bound, drawn where it exceeds one. Gray line: $3^w$.}
\label{fig:weight}
\end{figure*}

The size of the separation is set by the fingerprint's weight (Fig.~\ref{fig:weight}). For full weight it is
$(3/2)^r$, not $2^r$, because a GHZ-type probe covers a weight-$r$ Pauli with probability $\Theta((2/3)^r)$. Per-unit
stabilizer probes, GHZ-type and adaptive ones included, need \qnWrMeas{} times the vault's $\Nn$ at $r=10$, with a
$\log_2$ slope of \qnWrSlope{} in $r$. Against non-adaptive stabilizer probes the ratio is proved to be at least
\qnProvedWr{} (\qnProvedWrThm{} with the theorem's coverage). The adaptive GHZ heuristic reached
\qnAdaptiveOverLb{} times the non-adaptive bound, so it did not undercut it, but no general bound covers adaptive or
non-stabilizer probes. For $w=\lceil r/2\rceil$ the proved ratio is at least \qnProvedWhalf{} and the simulated one is
\qnWhalfMeas. The gain thus falls with the weight, because a lower weight is easier to cover with a single probe.

For low weight and for unscrambled tampering there is no exponential separation. For $w\le3$ the ratio stays below
$3^w$, and for $w\le2$ no separation is proved. Local tampering, which applies one fixed Pauli, gives a ratio of
\qnLocalRatio{}, flat in $r$; that is a constant-factor gain, and the pre-registered rule ``within $2\times$ means no
advantage'' was not met. A tool whose symmetry is local therefore gains little from a vault, and the exponential
regime needs a fingerprint that couples many cells.

\subsection{Results by attack}
\label{sec:res-attack}

\begin{table*}[t]
\caption{Results by attack and verifier class ($p_0=0.005$; 32-cell seal where the attack touches cells outside
$\Reg$). Vault results under extend, spray and mirror are conditional recoveries; the full report returns
\texttt{global\_anomaly} (extend, spray) or \texttt{copied\_errors} (mirror). $^\dagger$Conditional recovery, not a
success of the full report. The self-referencing column reports the implemented strategies, not the optimum of the
class.}
\label{tab:attacks}
\centering
\setlength{\tabcolsep}{3.5pt}
\renewcommand{\arraystretch}{1.15}
\begin{tabular}{L{2.0cm}L{4.0cm}L{4.1cm}L{3.1cm}L{3.3cm}}
\toprule
Attack & Per-unit & Self-referencing (P, P + sign) & In-seal reference & Vault \\
\midrule
Oblivious & Proposition~\ref{prop:chen}; \qnEoneRatioFine{}$\times$ the vault at $r=10$ & P + sign / vault
\qnPsignOverVaultFineLow{} ($\eps=0.15$), \qnPsignOverVaultFineHigh{} ($\eps=0.6$), simulated & Same as the vault:
\qnInsealObl$\times$ & $\Nn$ \qnVaultNninetyRtenLow{} / \qnVaultNninetyRtenHigh{} at $r=10$ \\
Extend & Chen setting even when told $\Reg$; proved $\ge$\qnProvedExtendPerUnit$\times$ the vault bound at $r=10$ &
$\Reg$ given: same law as oblivious. $\Reg$ unknown: sample bound \qnPsignBoundLow{} million units ($\eps=0.15$),
\qnPsignBoundHigh{} thousand ($\eps=0.6$), $r=4$--$10$ & References destroyed (\qnExtendInseal) & $^\dagger$$\Reg$ given:
\qnExtendGivenR$\times$ oblivious; $\Reg$ unknown: \qnSparseRatio$\times$ at $r=4$ (exact MAP), bound
\qnVaultBoundRtenLow{} / \qnVaultBoundRtenHigh{} units at $r=10$ \\
Spray & Product probe reads the copies: \qnProductRatio$\times$ the vault (interpolated \qnProductRatioInterp) & Not
run; copies stay inside a unit, so with $\Reg$ given P's law is unchanged & Blind for $\astar\notin C$
(\qnOracleNotInC); vault-like for $\astar\in C$ & $^\dagger$Unaffected without $\Reg$ (copies merged) \\
Mirror (knows the pairing) & Not run & This P scheme fails (at most \qnMirrorPaired{} per setting) & Blind
(\qnMirrorInseal{}) & $^\dagger$Unaffected: \qnMirrorVault$\times$ oblivious \\
\bottomrule
\end{tabular}
\end{table*}

Two kinds of success appear in Table~\ref{tab:attacks} and must not be confused. Under extend, spray and mirror the
vault entries are conditional recoveries: the decoder is told $\Reg$, searches a stated candidate space, or merges
copies with its own tier~1. The full report run on the same records stops at tier~1 with \texttt{global\_anomaly} or
\texttt{copied\_errors} and attributes nothing (Section~\ref{sec:claim-scope}). Full-pipeline attribution is established
only for the oblivious attacker.

\begin{figure*}[t]
\centering
\includegraphics[width=0.86\textwidth]{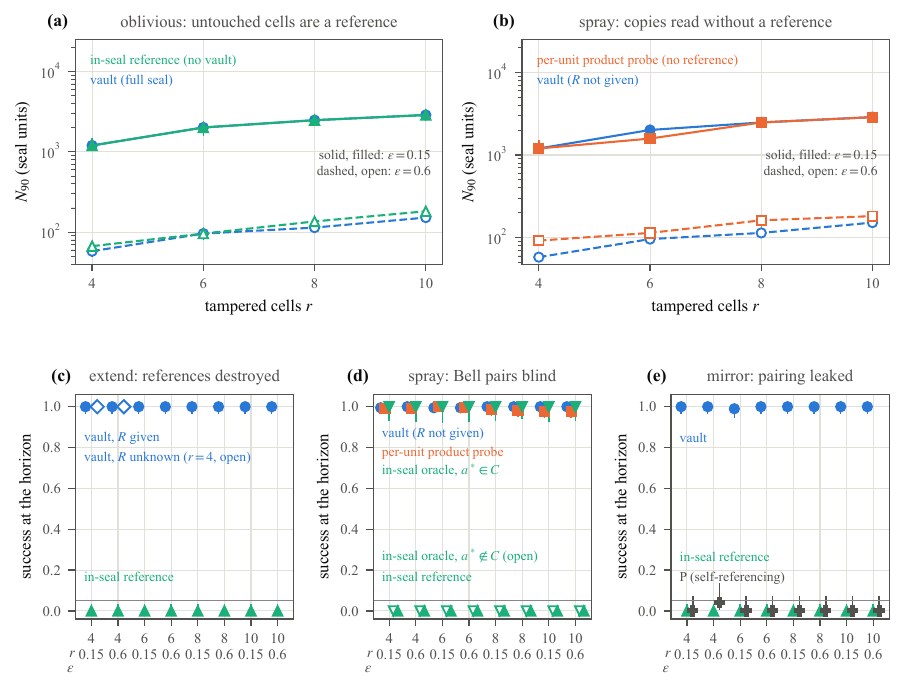}
\caption{Results by attack (32 cells, $p_0=0.005$). Colour is the verifier class, the marker the laboratory.
(a)~Oblivious attacker: $\Nn$ of the in-seal reference and of the full-seal vault, both including localization, with
90\% bootstrap intervals.
(b)~Spray attacker, who copies the error of tampered cell $i$ onto a group of cells outside $\Reg$ and needs no
secret: $\Nn$ of the vault, which is not given $\Reg$ and merges the copies in its tier~1, and of the per-unit product
probe with no reference, which reads one copy of each group in the $Z$ basis and another in the $X$ basis; same trials
and checkpoints ($\pm1$ grid step).
(c)--(e)~Success at the horizon with 95\% Clopper--Pearson intervals; the line at 0.05 is the pre-registered
``blind'' threshold.
(c)~Extend attacker: the vault is given $\Reg$ (no rate test can localize it); open diamonds are the vault with $\Reg$
unknown ($r=4$).
(d)~Spray (200 trials per laboratory): the in-seal oracle is told $\Reg$ and the groups and is split by whether
$\astar\in C$ (filled) or not (open).
(e)~Mirror attacker, who knows the pairing and copies each tampered cell's error onto its partner; P is mirrored
across its two units.}
\label{fig:attacks}
\end{figure*}

The value of a protected reference depends on the attack, and we do not merge the rows of Table~\ref{tab:attacks}
into one advantage. Each row has a simple reason, which the full-seal experiments of Fig.~\ref{fig:attacks} confirm,
one panel per attack.

\emph{Oblivious attacker: untouched cells are a reference.} A Bell pair between a tampered cell and an untouched cell
of the same seal reads the tampered cell's error times benign noise, which is exactly the vault's outcome law. An
in-seal verifier that first localizes $\Reg$ with secret random pairings and then pairs each tampered cell with an
untouched one reaches the vault's $\Nn$ within \qnInsealObl$\times$, including its localization cost
(Fig.~\ref{fig:attacks}(a)). Against an
oblivious attacker a vault is therefore not needed; its role is to \emph{guarantee} a reference that the attacker
cannot reach, which matters only against the attacks below. Proposition~\ref{prop:chen} accordingly concerns verifiers
without an untouched reference.

\emph{Extend: destroying the references.} An attacker who fully depolarizes every cell outside $\Reg$ needs no secret
and leaves no untouched reference, and in-seal references go blind (Fig.~\ref{fig:attacks}(c)). Every per-unit
verifier is then in the setting of Proposition~\ref{prop:chen},
even when told $\Reg$, because cells outside $\Reg$ carry no information (Lemma~\ref{lem:extend}); with $\Reg$ unknown,
the Fano bound over the vault's bound is at least \qnProvedExtendPerUnit{} at $r=10$. The vault is out of the
attacker's reach, so its law on $\Reg$ is unchanged: given $\Reg$ it costs \qnExtendGivenR$\times$ its oblivious cost
($\pm1$ grid step). Without $\Reg$, an exact search over all Paulis of weight at most four on the 32 cells costs
\qnSparseRatio$\times$ the vault with $\Reg$ given at $r=4$, and for larger $r$ only the bound of
Proposition~\ref{prop:ub} is available. P + sign keeps its law on $\Reg$ as well, and its sample bound with $\Reg$
unknown is polynomial in $r$ at zero noise or at a fixed effective bias. Destroying information is thus what makes the
protected reference necessary for per-unit verifiers.

\emph{Spray: copies blind Bell pairs but inform product probes.} An attacker who copies each tampered cell's error
onto a group of untouched cells needs no secret. Every in-seal Bell pair then reads a product of two error
components, which does not change when all components are multiplied by the same single-qubit Pauli
(Lemma~\ref{lem:spray}). The in-seal data are therefore independent of the fingerprint whenever $\astar$ does not
commute with those diagonal Paulis, about three quarters of the uniform family; an in-seal oracle told $\Reg$ and the
groups confirms this (\qnOracleNotInC{} at the horizon, Fig.~\ref{fig:attacks}(d)) and is vault-like for the rest
(\qnOracleInCRatio$\times$). But
the copies carry the error itself. A product probe with no reference at all reads one copy in the $Z$ basis and
another in the $X$ basis and so obtains each whole component (Lemma~\ref{lem:copies}). It learns the fingerprint at
\qnProductRatio$\times$ the vault's $\Nn$ (interpolated \qnProductRatioInterp; \qnProductSuccess{} at the horizon;
Fig.~\ref{fig:attacks}(b)).
Copying attackers are therefore a boundary of the method rather than a case for the vault. This conclusion concerns
the copying attackers we modelled, not all attackers.

\emph{Mirror: a leaked pairing.} An attacker who knows the pairing copies each tampered cell's error onto its partner,
so the pair reads only benign noise. In-seal references go blind (\qnMirrorInseal{} in every setting,
Fig.~\ref{fig:attacks}(e)), and when the
cross-unit pairing is known this P scheme fails as well (at most \qnMirrorPaired{} per setting, below the
pre-registered blind threshold of 0.05). The vault merges the copies and is unaffected (\qnMirrorVault$\times$
oblivious). Mirror requires a leaked secret, so it is a worst case rather than a typical threat.

\subsection{Tier 3: tool switches}
\label{sec:res-tier3}

\begin{figure*}[t]
\centering
\includegraphics[width=\textwidth]{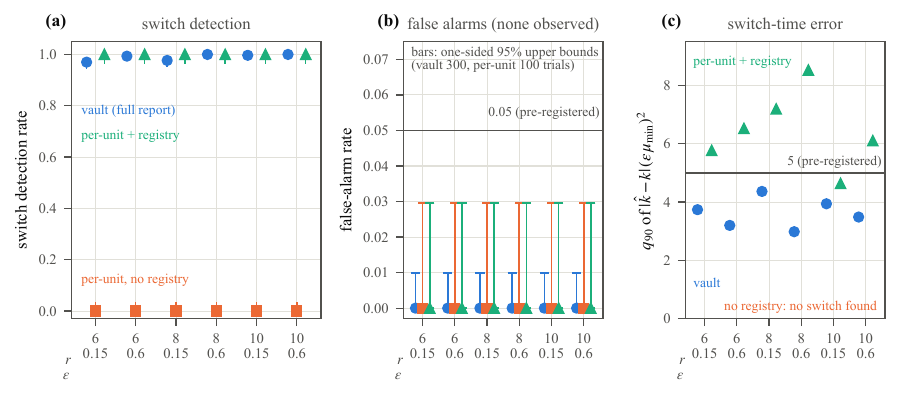}
\caption{Tier~3 on shared instances ($p_0=0.005$; the vault has 300 switch and 300 no-switch trials, each per-unit
detector 100 and 100).
(a)~Switch detection with 95\% Clopper--Pearson intervals.
(b)~False-alarm rate with one-sided 95\% upper bounds.
(c)~90\% quantile of the switch-time error, scaled by $(\eps\mu_{\min})^2$.
Lines are the pre-registered thresholds. These are simulated detectors for one switch between tools with different
$\astar$.}
\label{fig:tier3}
\end{figure*}

A switch between tools leaves the error rates unchanged, so only fingerprints can reveal it (Fig.~\ref{fig:tier3}).
At the vault's seal budget the full report detects at least \qnTierThreeVaultDetect{} of switches between tools with
different fingerprints, with \qnTierThreeVaultFA{} false alarms in every setting. A per-unit detector that must first
identify each window's fingerprint detects \qnTierThreeBlind{} at the same budget, because each window is far below
its identification threshold. A per-unit detector that already knows both fingerprints detects \qnTierThreeRegistry,
with a \qnTierThreeTimeErr$\times$ larger switch-time error, because testing a known fingerprint needs no memory.
These are results for these detectors, not a general impossibility: identifying a fingerprint and testing for a change
are different tasks. A registry of known fingerprints therefore removes the vault's advantage for tier~3, and the vault
matters for tools that have not been seen before.

\emph{Reordering in transit.} In a post-hoc test (E15) we regenerated E6's vault records and permuted a share of
their units, at random positions, before the report read them. With up to \qnReorderFstar{} of the units reordered
the report keeps E6's rule in all six settings (detection \qnReorderDetectFive, no false alarm). With 20\% reordered,
\qnReorderFailTwenty{} of the six settings fail, with up to \qnReorderFATwenty{} false switches at $\eps=0.15$, and a
full permutation leaves detection at \qnReorderDetectFull. All \qnReorderFATotal{} false switches came from fingerprints of negative sign.
Untampered units mixed in after the onset show the identity, whose parities are all $+1$, so they lift every parity by
their share, pull a negative fingerprint toward zero, and let each window pick a spurious positive parity.

\section{Engineering Requirements}
\label{sec:engineering}

This section turns the results into requirements for the scenario of Section~\ref{sec:scenario}. Every number is for
the simulated range ($r\le10$, $\eps\in\{0.15,0.6\}$, seal noise $p_0=0.005$ unless stated; E14 raises it to $0.1$);
Supplement~S8 gives the derivations.

\subsection{Units and vault qubits}
The attacker sets how many items it opens, so $\Nn$ is the number of items it can touch with one tool before the
verifier attributes the tool. Table~\ref{tab:inventory} gives this threshold for the vault and for the per-unit
strategy, together with the vault's inventory. When a whole consignment is issued at once and all its modules are
outstanding together, the vault holds $n\Nn$ qubits, here for an $n=32$-cell module, and must keep each of them coherent
for the round-trip time $T$ of its module. At $r=10$ and $\eps=0.15$ this is \qnVaultQubitsRtenLow{} vault qubits,
against \qnVaultQubitsRtenHigh{} at $\eps=0.6$. Over the simulated range the vault's $\Nn$ is
(\qnVaultConst)$\,r/\eps^2$. If the attacker touches only a fraction $f$ of the consignment, the consignment must be
$\Nn/f$ items long, and the inventory grows by $1/f$ (for a decoder that uses only the tampered units;
Section~\ref{sec:limitations}). If modules are issued continuously, the inventory is $n$ times
the number of modules in transit, the issue rate times $T$, and attribution waits until $\Nn$ tampered modules have
arrived.

\begin{table}[t]
\caption{Attribution threshold and vault inventory ($p_0=0.005$, fine checkpoints). Units: tampered items needed for
90\% identification of $(\astar,s)$. Vault qubits: $32\Nn$, held for the round-trip time $T$ when the units are
outstanding together.}
\label{tab:inventory}
\centering
\setlength{\tabcolsep}{3pt}
\begin{tabular}{rrrrrr}
\toprule
$r$ & $\eps$ & vault $\Nn$ & vault qubits & per-unit $\Nn$ & ratio \\
\midrule
\qnResourceRows
\bottomrule
\end{tabular}
\end{table}

\subsection{Coherence of vault and seal}

\begin{figure*}[t]
\centering
\includegraphics[width=0.78\textwidth]{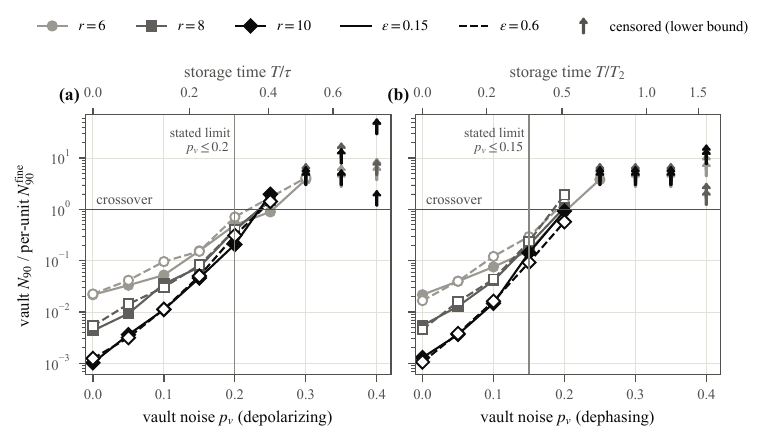}
\caption{Vault noise. The $y$ axis is the vault's $\Nn$ over the per-unit $\Nn$ at the same $(r,\eps)$, the latter
on fine checkpoints; below 1 the vault is ahead. Arrows mark censored vault runs (lower bounds). The vertical line is
the stated limit. The top axis converts $p_v$ to a storage time under exponential decay, Eq.~(\ref{eq:lifetime}). The
vault side is read on the grid, which comes late, so the crossover is on the low side.}
\label{fig:noise}
\end{figure*}

Fig.~\ref{fig:noise} adds depolarizing or dephasing noise $p_v$ to the vault half only. The vault stays ahead of the
per-unit strategy in every tested setting up to $p_v=0.2$ (depolarizing) and $p_v=0.15$ (dephasing); the crossovers lie
at \qnCrossDep{} and \qnCrossDeph. These are finite-size results: at a fixed per-qubit noise the vault's cost grows
with the fingerprint weight as $\eps_{\mathrm{eff}}^{-2}$, so we do not extrapolate them to larger $r$.

Under an exponential decay model the limits become storage times. For a memory with dephasing time $T_2$ the flip
probability after a time $T$ is $\tfrac12(1-e^{-T/T_2})$, and for depolarization with time constant $\tau$ it is
$\tfrac34(1-e^{-T/\tau})$, which gives
\begin{equation}
T\le\qnLifeVaultDeph\,T_2,\qquad T\le\qnLifeVaultDep\,\tau
\label{eq:lifetime}
\end{equation}
for the vault. The seal cells have a much smaller budget. A seal cell of coherence time $\tau_s$ held for a time $T$
reaches the benign noise $p_0$ with $1-4p_0/3=e^{-T/\tau_s}$, so
\begin{equation}
T\le-\tau_s\ln\!\bigl(1-4p_0/3\bigr),
\label{eq:seal-lifetime}
\end{equation}
and the same for a dephasing seal matched on the attenuated eigenvalue. At $p_0=0.005$, the noise assumed in the main
experiments, this is $T\le\qnLifeSeal\,\tau_s$: if seal and vault are built from the same kind of qubit, the seal
cells must stay coherent for about 150 times the round trip.

\begin{figure*}[t]
\centering
\includegraphics[width=\textwidth]{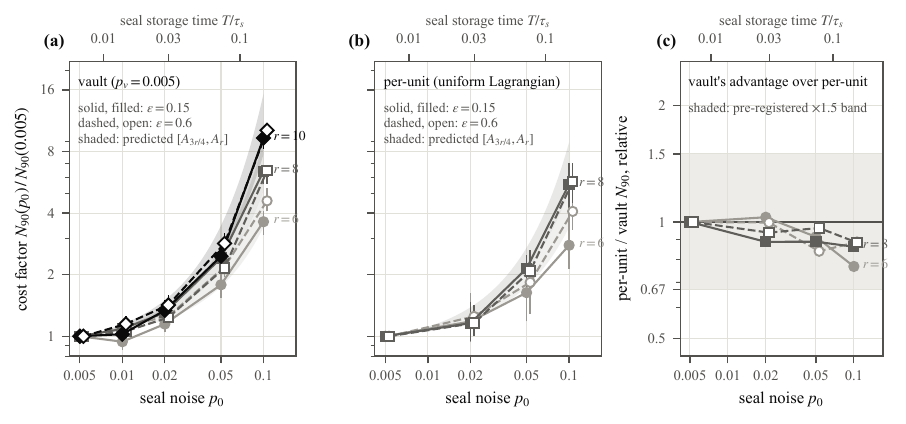}
\caption{Seal requirement curve (E14; vault noise fixed at $p_v=0.005$; E1 instances; vault 200 trials, per-unit
50). (a), (b)~Attribution cost at seal noise $p_0$ relative to $p_0=0.005$, $\rho=\Nn(p_0)/\Nn(0.005)$, for the vault
and the per-unit strategy, with paired 90\% bootstrap intervals; shaded: the pre-registered band
$[A_{3r/4},A_r]$ of the seal attenuation law. (c)~The per-unit over the vault $\Nn$, relative to its value at
$p_0=0.005$; shaded: the pre-registered $\times1.5$ band. Top axes: the seal storage time $T/\tau_s$ of
Eq.~(\ref{eq:seal-lifetime}). Every trial of a group shares the group's checkpoints, so $\Nn$ is read without the
late-grid effect.}
\label{fig:seal}
\end{figure*}

How much seal noise attribution tolerates is a trade-off, and E14 simulates it (Fig.~\ref{fig:seal}). Holding the
vault's own noise at $0.005$, we raised the seal noise to $p_0=0.1$ on E1's instances; the protocol, the predicted band
and the decision rules were fixed before the run (Supplement~S1). Both laboratories read the seal half once, so the
fingerprint is attenuated by $(1-4p_0/3)^{|\astar|}$ for both, and $\Nn$ should grow by
$A_w(p_0)=\bigl((1-4\cdot0.005/3)/(1-4p_0/3)\bigr)^{2w}$ with $w$ between $3r/4$ and $r$. The simulated cost factor
overlaps this band in \qnSealPOne{} groups; the exception, the vault at $r=6$, $\eps=0.15$ and $p_0=0.01$, gave
\qnSealFailRho{} against \qnSealFailBand. At $r=10$ the vault's cost grows by \qnSealRhoTwo{} at $p_0=0.02$,
\qnSealRhoFive{} at $p_0=0.05$ and \qnSealRhoTen{} at $p_0=0.1$. Because the seal noise acts on both laboratories
alike, the per-unit over the vault $\Nn$ changes only by a factor \qnSealRatioChange{} (\qnSealPTwo{} groups inside the
pre-registered $\times1.5$ band), and the controls at $p_0=0.005$ reproduce E1f (\qnSealPThree).

The seal requirement is therefore a curve, not a limit. A seal cell that may reach $p_0=0.05$ can be held for
$\qnSealLifeFive\,\tau_s$, \qnSealLifeGainFive{} times as long as at $p_0=0.005$, at \qnSealRhoFive{} times the
attribution cost; one that may reach $p_0=0.1$ is held for $\qnSealLifeTen\,\tau_s$ at about ten times the cost. These
are finite-size results for $r\le10$, and the cost factor grows with the fingerprint weight. Tier~1 and the full report
were not evaluated above $p_0=0.005$.

Table~\ref{tab:resources} applies (\ref{eq:lifetime}) and (\ref{eq:seal-lifetime}) to reported coherence times. The
entries are laboratory records for single qubits, not for modules in transit, and we read each reported coherence time
as a dephasing time $T_2$ of an exponential decay. Even with the longest record listed, a seal cell held to $p_0=0.005$
lasts minutes and one allowed $p_0=0.05$ about half an hour, while the vault could hold for hours. Of the two lifetimes
the seal's is therefore the binding one, and seals that survive a logistics round trip of days remain the main lifetime
gap: tolerating more seal noise buys time at a simulated cost, but not days.

\begin{table*}[t]
\caption{Reported coherence and Bell-state figures, and the storage times they would allow under
(\ref{eq:lifetime}) and (\ref{eq:seal-lifetime}). Values are quoted from the cited works; blank cells were not
filled because we found no source we could check. Every listed value is a coherence time measured with dephasing or
dynamical-decoupling sequences, so every row reads it as a dephasing time $T_2$ and uses the dephasing forms: vault
$T=\qnLifeVaultDeph\,T_2$ from (\ref{eq:lifetime}), and seal $T=\qnLifeSeal\,T_2$ at $p_0=0.005$ and
$T=\qnSealLifeFive\,T_2$ at $p_0=0.05$ from (\ref{eq:seal-lifetime}). Storage times are orders of magnitude for one
qubit in a laboratory.}
\label{tab:resources}
\centering
\setlength{\tabcolsep}{4pt}
\begin{tabular}{L{2.65cm}L{3.65cm}L{1.6cm}L{1.6cm}L{1.6cm}L{4.3cm}}
\toprule
Platform & Reported memory coherence & Vault $T$ & Seal $T$, $p_0=0.005$ & Seal $T$, $p_0=0.05$ & Bell-state or
Bell-measurement figure \\
\midrule
Trapped ion & single ${}^{171}$Yb$^+$ ion qubit, estimated over one hour (exponential fit to data up to 960 s;
read here as $T_2=1$ h) \cite{wang2021single} & $\sim20$ min & $\sim20$ s & $\sim4$ min &
${}^{40}$Ca$^+$ Bell-state preparation infidelity $6(3)\times10^{-4}$ \cite{clark2021high} \\
NV centre & spin register in diamond, coherence up to 63(2) s \cite{bradley2019ten} & $\sim20$ s & $\sim0.4$ s &
$\sim4$ s & \\
Rare-earth ion & ${}^{151}$Eu$^{3+}$:Y$_2$SiO$_5$ nuclear spin, $370\pm60$ min with dynamic decoupling at 2 K
\cite{zhong2015optically} & $\sim2$ h & $\sim2$ min & $\sim25$ min & optical interface: a linear-optical Bell-state
analyzer succeeds with probability at most $1/2$ \cite{calsamiglia2001maximum} \\
Superconducting & transmon, coherence with dynamical decoupling above 0.3 ms \cite{place2021new} & $\sim0.1$ ms &
$\sim2$ $\mu$s & $\sim20$ $\mu$s & \\
\bottomrule
\end{tabular}
\end{table*}

\subsection{Operations}
\emph{Issue.} Each cell needs one Bell pair and a uniformly random Pauli frame, two classical bits per cell and unit.

\emph{Verification and measurement errors.} The verifier undoes the frame and performs one Bell measurement per
cell, so the quantum depth is constant in $r$. The simulations treat the Bell measurement as ideal. An imperfect Bell
measurement modelled as two-qubit depolarizing noise of rate $p_{\mathrm{BM}}$ on the pair just before an ideal
measurement changes the outcome by the product of the two Paulis. Of the 15 non-identity two-qubit Paulis, three have
an identity product and twelve give $X$, $Y$ or $Z$ equally often, so the outcome suffers single-qubit depolarizing
noise with non-identity probability $\tfrac45p_{\mathrm{BM}}$. That is noise on the vault half, and its attenuation
multiplies that of the memory, so the vault-noise limit above bounds memory and measurement noise together. Errors that
are correlated across cells are not covered.

\emph{Frame secrecy.} The frame is a one-time classical secret, 64 bits for a 32-cell module, that must stay
confidential and intact from issue to verification, with the same handling as a cryptographic key. A leaked frame
lets the attacker act coherently instead of being twirled (the coherent check of Section~\ref{sec:model} deviates by
\qnCircuitNoFrame{} without the frame), and a leaked pairing enables mirror.

\emph{Classical decoding.} The exact posterior over $4^r$ hypotheses, or the Walsh--Hadamard transform in the report,
is practical up to $r=12$; beyond that the report returns \texttt{not\_decoded}.

\emph{Preparation timing.} Strategies that adapt the preparation of later units need units issued after earlier
reads, which in the consignment scenario means a second consignment. Measurement adaptivity is always available.

\section{Boundaries and Limitations}
\label{sec:limitations}

\subsection{Where the vault gives no exponential gain}
\begin{itemize}
\item \emph{Low weight and local tampering.} Fingerprints of weight at most three and unscrambled tampering show
constant-factor gains that do not grow with $r$ (Section~\ref{sec:res-weight}).
\item \emph{Known fingerprints.} A known fingerprint can be tested without memory, so a registry removes the tier-3
advantage (Section~\ref{sec:res-tier3}).
\item \emph{Copying attackers.} Spray copies errors onto untouched cells and so exposes them to reference-free
product probes, which then match the vault (Section~\ref{sec:res-attack}).
\item \emph{Self-referencing verifiers.} Against the implemented P + sign the gain is a simulated factor of about
$2/\eps^2$. We have no lower bound for the self-referencing class as a whole.
\item \emph{In-seal references.} Against an oblivious attacker, untouched cells of the same seal do as well as the
vault.
\item \emph{Attackers who erase the fingerprint.} A uniformly random Pauli on the tampered cells, drawn afresh for
every unit, removes the fingerprint exactly and costs the attacker only single-qubit gates and random bits; a biased
layer leaves a residual $\eps\beta$ and raises the attribution cost about $\beta^{-2}$-fold. Forensics concerns
attackers who do not erase their fingerprint. Against the others only tier~1 works (Section~\ref{sec:model}).
\end{itemize}

\subsection{Scope of the claims}
\label{sec:claim-scope}
Detection and localization (tier~1) need no fingerprint; attribution (tiers~2 and~3) needs one that the attacker has
not erased and, in the full report, an oblivious attacker. Table~\ref{tab:capability} sets out what the full report
delivers against each attacker.

\begin{table}[t]
\caption{What the full report (Algorithm~\ref{alg:qmf}) delivers, by attacker capability. ``Conditional'' means a
separate decoder told $\Reg$ or a candidate space (Table~\ref{tab:attacks}). The erasing row follows from
Section~\ref{sec:tamper}: for $w\ge2$ the cells' error rates do not depend on $\eps$, and for $w=1$ one cell's rate
moves by at most $\eps/4$ and stays far above the benign rate; the other rows from
Sections~\ref{sec:report} and~\ref{sec:results} and the unit tests. Rows assume that the tampered units follow the
onset contiguously in arrival order (Section~\ref{sec:claim-scope}).}
\label{tab:capability}
\centering
\setlength{\tabcolsep}{3pt}
\begin{tabular}{L{1.53cm}L{2.96cm}L{1.7cm}L{1.5cm}}
\toprule
Attacker & Tier 1: detect, localize & Tier 2: fingerprint & Tier 3: switch \\
\midrule
Oblivious & localized (\texttt{localized}) & decoded & one switch between different $\astar$ \\
Erasing (random Pauli layer) & localized; rates stay high & nothing to decode, for any verifier & nothing to detect \\
Extend & detected, not localized (\texttt{global\_anomaly}) & conditional only & not evaluated \\
Spray & detected, not localized (\texttt{global\_anomaly}) & conditional only & not evaluated \\
Mirror & cells and copies flagged (\texttt{copied\_errors}) & conditional only & not evaluated \\
\bottomrule
\end{tabular}
\end{table}

Proposition~\ref{prop:chen} needs $\eps\le1/3$, one parity, the per-unit class and the tool's own channel. It is an
asymptotic statement: with its explicit constant, (\ref{eq:chen-explicit}), it proves a separation against every
per-unit verifier only from $r=\qnChenRstar$ at $\eps=0.15$ and $p_0=0.005$, beyond our simulated range.
Propositions~\ref{prop:fano} and~\ref{prop:ub}, and the proved ratios of Table~\ref{tab:proved}, concern non-adaptive
stabilizer probes, and no general bound covers adaptive or non-stabilizer per-unit probes. The seal noise is
$p_0\le0.005$ in the main experiments and up to $0.1$ in E14, which simulated fingerprint recovery only; tier~1 and the
full report were not evaluated above $p_0=0.005$. The simulations cover
$r\le10$, and the full-seal experiments use 32 cells. With fixed per-qubit noise the effective bias decays
exponentially with the weight, so the noisy separations and the vault-noise limits are finite-size results. The
vault's cost with $\Reg$ unknown was simulated at $r=4$ only; above that, and for P + sign, only information-theoretic
sample bounds are given, which say nothing about decoding time.

Tool-switch detection is evaluated in simulation for one switch between tools with different fingerprints, in records whose arrival
order keeps the attack order; it tolerated reordering of up to \qnReorderFstar{} of the units (E15). More
generally, the full report assumes that every unit after the onset is tampered. Untampered units interleaved with
tampered ones, by reordering or by an attacker who touches only part of a consignment, lift every parity, and the
report would then need a decoder that drops units without an error on $\hat\Reg$, which we did not implement. Sign
changes with the
same fingerprint are flagged as unsupported, and changes of strength alone or repeated changes are neither detected nor
evaluated. Incident linking is implemented and unit-tested but not evaluated on populations of incidents.
Attribution is up to fingerprint equivalence (Section~\ref{sec:tamper}): tools with the same fingerprint, and an
attacker who reproduces another tool's fingerprint, are not told apart. All
attackers use the same channel on every unit and do not know the Pauli frame; extend and spray need no secret, and
mirror needs the pairing. Results for copying attackers concern the attackers we modelled.

Each of the report's tests has its own decision rule (Supplement~S4); only the tier-2 threshold admits a theoretical
error bound, conditional on the region and onset, the whole report has no established family-wise error guarantee, and
the \qnTierThreeVaultFA{} false alarms of tier~3 are a simulation result for the E6 scenario, not a guaranteed rate.
$\Nn$ is an empirical crossing, some groups have 20 trials, and grid values from groups with varying horizons are
$\pm1$ grid step. There is no hardware run: the circuits were checked against the sampler in exact simulation only,
the Bell measurement is ideal in the simulations, and the platform figures of Table~\ref{tab:resources} are laboratory
records for single qubits. Every verifier uses quantum probes, so the comparison is about sample complexity between
resource classes, not about classical against quantum computation.

\section{Conclusion}
\label{sec:conclusion}

Quantum seals can do more than detect tampering: with a protected reference they can attribute it. QMF localizes the
tampering, learns the hidden-parity fingerprint of the tool and detects a switch between tools in arrival order, and it says
explicitly
when an attack is outside its supported scope.

Whether the protected reference is worth its cost depends on the allowed resources and on the attack. In the specified
per-unit resource model the sample separation is asymptotically exponential, by reduction for one parity of strength
at most $1/3$. At the simulated sizes it is proved as a ratio against non-adaptive stabilizer probes and observed in simulation
to be larger. Against the implemented
self-referencing alternative the difference is a simulated polynomial factor of about $2/\eps^2$, and we claim no
optimality over the whole self-referencing class. Untouched cells of the seal replace the vault against an oblivious
attacker; an attacker who destroys them makes the vault necessary for per-unit verifiers, and an attacker who copies
errors gives the fingerprint away to verifiers without any reference.

A vault is therefore worth deploying when items must be verified one at a time, when the tool scrambles a fingerprint
of high weight, and when attackers may destroy the untouched cells. Against such an attacker the present full report
only flags a global anomaly; the vault keeps the fingerprint, but recovering it needs the tampered region from another
source or a sparse search over the seal, which we ran at $r=4$ only. Forensics concerns attackers who do not erase
their fingerprint: an attacker who knows QMF can erase it with a random Pauli layer at little cost, and then no
reference helps beyond tier~1. Under independent noise the binding lifetime is the seal's, not the vault's, which
tolerates substantial memory noise in the simulated range; a vault of thousands to tens of thousands of qubits, with its
readout and correlated noise, is itself untested. The seal's noise budget is a simulated trade-off: a seal allowed five percent noise lasts
\qnSealLifeGainFive{} times longer at \qnSealRhoFive{} times the attribution cost, yet even then current laboratory
memories reach tens of minutes, not the days of a logistics round trip. Three questions remain open: a lower bound for the whole
self-referencing class, bounds for adaptive and non-stabilizer per-unit probes, and a hardware demonstration with a
real quantum memory and seals that survive transport.

\section*{Data Availability}
{\sloppy %
The code, raw results, analysis and the scripts that rebuild every result figure and table and every headline number
are archived on Zenodo (doi:~10.5281/zenodo.24148790), with the SHA-256 hash of every file in \texttt{SHA256SUMS}, and
are available at
\url{https://github.com/Jeong-Jaeyun/Quantum-Memory-Forensics-for-Tampered-Quantum-Seals-Resource-Dependent-Sample-Gains}.
The supplement contains the experiment register, further results, the circuit checks and the reproduction
procedure.\par}

\appendix
\section{Proofs}
\label{app:proofs}

Throughout, Paulis on $k$ qubits are identified with $\mathbb F_2^{2k}$ up to phase, so products of Paulis are sums
of vectors, and $\chi_a(E)=(-1)^{\langle a,E\rangle}$. A Pauli channel with eigenvalues $\lambda_b$ has error law
$p(E)=4^{-k}\sum_b\lambda_b\chi_b(E)$ and satisfies $\mathrm{Tr}[b\,\Lambda(\rho)]=\lambda_b\mathrm{Tr}[b\rho]$.

\subsection{Proof of the per-unit lower bound (Proposition~1)}
\emph{Noise.} The per-unit verifier sees the seal half only, so the channel it probes is $\Lambda'=D\circ\Lambda\circ
D'$, where $D,D'$ are products of single-qubit depolarizing channels. A verifier given the noiseless $\Lambda$ can
apply $D'$ to its probe before the attack and $D$ after it, using random Paulis and classical randomness only. Any
per-unit protocol for $\Lambda'$ with $N$ units is therefore a per-unit protocol for $\Lambda$ with $N$ units, and a
lower bound at $p_0=0$ holds for every $p_0$.

\emph{The game.} We use the game of Ref.~\cite{chen2024tight} (Supplemental Material, Sec.~S3) with $n=r$ and
$\eps_0=\eps\le1/3$. A referee draws $a\neq I$ and $s=\pm1$ uniformly. With probability $1/2$ it lets the player use
$\Lambda_0$, otherwise $\Lambda_{a,s}$, whose only non-identity Pauli eigenvalue is $s\eps$ at $a$; this is the
channel (\ref{eq:scramble}) with $(\astar,s)=(a,s)$. The player must finish all quantum measurements before $a$ is
revealed, and wins if it names the hypothesis.

\emph{The player.} Let $\mathcal A$ be a per-unit verifier that outputs $(\hat a,\hat s)$ after $N$ units and is
correct with probability at least $0.9$ averaged over the uniform prior on $(a,s)$. Since $a\neq I$, an output
$\hat a=I$ is never correct, so we may assume $\hat a\neq I$ (replace $I$ by any fixed non-identity Pauli). The player
runs $\mathcal A$ on $N$ channel uses. It then prepares $m=\lceil8\ln20/\eps^2\rceil$ further probes, each a product of single-qubit
eigenstates of the local factors of $\hat a$ with $\hat a$-eigenvalue $+1$, applies the channel, measures each qubit in
the local basis of $\hat a$, and records the product $Y_j\in\{\pm1\}$ of the outcomes. Then $\mathbb E[Y_j]=\lambda_{\hat a}$.
The player declares $\Lambda_{a,\pm}$ if $\hat s\bar Y>\eps/2$, with $\bar Y$ the mean of the $Y_j$, and $\Lambda_0$
otherwise. Under $\Lambda_0$, $\lambda_{\hat a}=0$, so Hoeffding's inequality \cite{hoeffding1963probability} gives
an error probability at most $e^{-m\eps^2/8}\le1/20$, whatever $\mathcal A$ output. Under $\Lambda_{a,s}$ with
$(\hat a,\hat s)=(a,s)$, $\mathbb E[\hat s\bar Y]=\eps$ and the error probability is again at most $1/20$. The player
therefore wins with probability at least $\tfrac12\cdot0.95+\tfrac12\cdot0.9\cdot0.95>0.90$, averaged over the
referee's randomness. Every step uses per-unit resources only.

\emph{The bound.} For each $a$, the probability of winning given $a$ is at most $\tfrac12+\tfrac12\,
\mathrm{TVD}(p_0,\mathbb E_s p_{a,s})$, where $p_0$ and $p_{a,s}$ are the laws of the player's whole transcript.
Averaging over $a$ gives $\mathbb E_a\mathrm{TVD}(p_0,\mathbb E_s p_{a,s})\ge0.8>1/3$, which is condition (S15) of
Ref.~\cite{chen2024tight}. The remainder of the proof of their Theorem~2 bounds this averaged distance for every
entanglement-free protocol with classical memory and adaptivity that uses the channel $N'$ times. Their Eq.~(S69)
reads $\mathbb E_a\mathrm{TVD}\le N'\eps_0^2\,\tfrac{2^n}{4^n-1}\bigl(1+2\sqrt{f(\eps_0)}\bigr)$, with
\begin{equation*}
f(\eps_0)=\frac12\Bigl[\Bigl(\frac{2}{1-\eps_0^2}\Bigr)^2+\frac{8}{(1-\eps_0)^2(1-2\eps_0-\eps_0^2)}\Bigr]
\end{equation*}
from their Eq.~(S67), which is finite and below $44$ for $\eps_0\le1/3$. With $N'=N+m$, $n=r$, $\eps_0=\eps$ and
$\mathbb E_a\mathrm{TVD}\ge0.8$ this gives (\ref{eq:chen-explicit}). Hence $N+m=\Omega(2^r/\eps^2)$, and since
$m=O(1/\eps^2)$, $N=\Omega(2^r/\eps^2)$. \hfill$\blacksquare$

\subsection{Proof of the Fano bound (Proposition~2)}
Let $X=(\astar,s)$ be uniform over $W_w\times\{\pm1\}$, so $\log_2|X|=\log_2(2M)$, and let $Y_u$ be the $r$ syndrome
bits of unit $u$, whose stabilizer group $S_u$ is fixed before any data (non-adaptive). Given $X$, a syndrome bit of
$g\in S_u$ has mean $s\eps'$ if $g=\astar$ and is otherwise uniform and independent of the others, because the
attacked channel has a single non-identity eigenvalue. Hence $H(Y_u\mid X)=r-d(\eps')$ if $\astar\in S_u$ and $r$
otherwise, while $H(Y_u)\le r$. So $I(X;Y_u)\le\Pr[\astar\in S_u]\,d(\eps')\le\ceff(r,w)\,d(\eps')$, because $S_u$ is
a Lagrangian and $\astar$ is uniform over $W_w$. Given $X$ the units are independent, so
$I(X;Y_1,\dots,Y_N)\le\sum_uI(X;Y_u)\le N\ceff d(\eps')$. Fano's inequality \cite{cover2006elements} requires
$I\ge0.9\log_2(2M)-1$ for a success probability of $0.9$ averaged over the prior, which gives (\ref{eq:fano}). A
verifier that does not know $\Reg$ is in particular a verifier that ignores the knowledge of $\Reg$, so the bound also
holds when $\Reg$ is unknown. \hfill$\blacksquare$

\subsection{Proof of the upper bounds (Proposition~3)}
For $a\neq b$ the characters $\chi_a$ and $\chi_b$ are independent uniform signs under the uniform law, so
$\mathrm{BC}=\tfrac14\sum_{x,y=\pm1}\sqrt{(1+s\eps_ax)(1+s'\eps_by)}=g(\eps_a)g(\eps_b)$; for $a=b$ and $s\neq s'$,
$\mathrm{BC}=\tfrac12\sum_{x=\pm1}\sqrt{(1+\eps_ax)(1-\eps_ax)}=\sqrt{1-\eps_a^2}$. Both are at most
$1-\eps_{\min}^2/8$, using $g(x)\le1-x^2/8$ and $\sqrt{1-x^2}\le1-x^2/2$. Maximum likelihood over $N$ independent
samples confuses $h$ with $h'$ with probability at most $\mathrm{BC}(P_h,P_{h'})^N$, so the union bound gives an error
of at most $|H|e^{-N\eps_{\min}^2/8}\le\delta$ once $N\ge8\ln(|H|/\delta)/\eps_{\min}^2$.

For the vault, a Pauli $a$ of weight at most $r$ is attenuated by $(1-4p_0/3)^{2|a|}\ge(1-4p_0/3)^{2r}$ (seal and vault
halves), which gives $\eps_{\min}$, and $|H|=2(4^r-1)$ with $\Reg$ given or $2M_n$ with $\Reg$ unknown. Since
$\binom nk3^k$ increases in $k$ for $k\le r\le3n/4$, $M_n\le r\binom nr3^r\le r(3en/r)^r$. For P + sign, a pair of
units reveals $E_uE_{u'}$, whose law is (\ref{eq:scramble}) with strength $\eps^2$ and no sign, attenuated by the two
seal halves; with $M_n$ sign-free hypotheses the pair part needs $8\ln(M_n/\delta_a)/\eps_P^2$ pairs. Given the right
$\hat a$, the sign phase measures the eigenvalue of $\hat a$ on single units, with mean $s\eps_s$ or larger in modulus,
and Hoeffding's inequality makes the sign of the average wrong with probability at most $e^{-m\eps_s^2/2}\le\delta_s$.
The errors add. \hfill$\blacksquare$

\subsection{Lemmas for the attacks}
\begin{lemma}[Extend]
\label{lem:extend}
Under extend, the syndromes of a per-unit stabilizer probe with group $S_u$ depend on $(\astar,s)$ only through the
parities of $S_u\cap\mathcal P_\Reg$, the elements supported on $\Reg$. This is an isotropic subgroup of
$\mathcal P_\Reg$, so the probe covers a weight-$w$ fingerprint with probability at most $\ceff(r,w)$.
\end{lemma}
\begin{proof}
Write $E=(E_\Reg,E_{\bar\Reg})$. Under extend $E_{\bar\Reg}$ is uniform and independent of $E_\Reg$. For $g\in S_u$
the syndrome is $\langle g_\Reg,E_\Reg\rangle+\langle g_{\bar\Reg},E_{\bar\Reg}\rangle$. The map
$g\mapsto\langle g_{\bar\Reg},E_{\bar\Reg}\rangle$ is a uniformly random linear functional on the restriction of
$S_u$ to $\bar\Reg$, independent of everything else, so it randomizes every syndrome combination except those of
elements with $g_{\bar\Reg}=I$. The remaining elements form $S_u\cap\mathcal P_\Reg$, which is isotropic because $S_u$
is, and every isotropic subgroup lies in a Lagrangian of $\mathcal P_\Reg$.
\end{proof}

\begin{lemma}[Spray and in-seal Bell pairs]
\label{lem:spray}
Under spray every cell carries one component $E_i$ of $E=(E_1,\dots,E_r)$, plus benign noise. Let
$D=\{P^{\otimes r}:P\in\{I,X,Y,Z\}\}$ and $C=\{a:\sum_ia_i=0\}$, the Paulis whose cell codes add to zero.
(i) Every in-seal Bell-pair outcome is a function of $E$ modulo $D$, whatever the pairing.
(ii) The law of $E$ modulo $D$ is uniform unless $\astar\in C$. A fraction $(4^{r-1}-1)/(4^r-1)$, about one quarter,
of the uniform family lies in $C$.
\end{lemma}
\begin{proof}
(i) A Bell pair of two seal cells carrying components $i$ and $k$ reads their product $E_i+E_k$, which is unchanged
when every component is shifted by the same single-qubit Pauli. (ii) The law of the coset $E+D$ is
$\sum_{d\in D}p(E+d)=4^{-r}\bigl(4+s\eps\,\chi_{\astar}(E)\sum_{d\in D}\chi_{\astar}(d)\bigr)$. For $d=P^{\otimes r}$,
$\langle\astar,d\rangle=\langle\sum_i\astar_i,P\rangle$, and the character sum over $P$ is $4$ if
$\sum_i\astar_i=0$ and $0$ otherwise. Exactly $4^{r-1}$ vectors $a$ have $\sum_ia_i=0$, one of them the identity.
\end{proof}

\begin{lemma}[Copies in complementary bases]
\label{lem:copies}
If two cells $j\neq j'$ both carry the component $E_i$, a probe that prepares $j$ in $|0\rangle$ and $j'$ in
$|+\rangle$ and measures $Z$ on $j$ and $X$ on $j'$ reads the $x$ and $z$ bits of $E_i$, each flipped independently
with probability $2p_0/3$ by benign noise. One unit then gives a sample of $E$ on the $r$ components, with parity $b$
attenuated by $(1-4p_0/3)^{\mathrm{bits}(b)}$, where $\mathrm{bits}(b)$ counts the $x$ and $z$ bits of $b$.
\end{lemma}
\begin{proof}
A $Z$ measurement of $E|0\rangle$ flips exactly when $E$ anticommutes with $Z$, that is, when its $x$ bit is one;
likewise for $X$ on $|+\rangle$ and the $z$ bit. Depolarizing noise flips a given measured bit with probability $2p_0/3$,
since two of the three non-identity Paulis anticommute with the measured one. The frame is undone before the
measurement, as for every verifier.
\end{proof}
Spray's groups hold $(n-r)/r$ cells, at least two for $r\le n/3$, so the lemma applies at every $r\le10$ of our 32-cell
seal.

\section{Physical realization of the tamper model}
\label{app:physical}
\emph{Construction for any $r$.} Let $\mathcal C(\astar)$ be the centralizer of $\astar$, the Paulis that commute with
it. As a subspace of $\mathbb F_2^{2r}$ it has dimension $2r-1$ (index two) and contains $\astar$, so $2r-1$
generators span it, and its symplectic complement is $\{I,\astar\}$. Applying a uniformly random element of $\mathcal C(\astar)$ multiplies each Pauli eigenvalue
$\lambda_b$ by $\mathbb E_g\chi_b(g)$, which is $1$ for $b\in\{I,\astar\}$ and $0$ otherwise. Starting from the
identity channel, this leaves $\lambda_I=\lambda_{\astar}=1$: a tool whose dynamics have the symmetry $\astar$
scrambles everything except $\astar$. Applying afterwards a Pauli $Q$ that anticommutes with $\astar$, with probability
$p$, multiplies $\lambda_{\astar}$ by $1-2p$. With $p=(1-s\eps)/2$ the result is (\ref{eq:scramble}). A circuit
therefore needs $2r$ tool qubits: $2r-1$ prepared in $|+\rangle$, each controlling one generator of
$\mathcal C(\astar)$, and one prepared in $R_Y(\theta)|0\rangle$ with $\sin^2(\theta/2)=(1-s\eps)/2$, controlling $Q$.
Discarding the tool qubits leaves the channel on the cells.

\emph{The $r=3$ circuit and its checks.} For $\astar=X\otimes Y\otimes Z$ the five generators are $X_0$, $Y_1$, $Z_2$,
$Z_0Z_1$ and $Z_1X_2$, and $Q=Z_0$. Exact Qiskit simulation of this circuit, together with the issue, probe and
verification blocks, reproduces the sampler's laws in \qnCircuitLawChecks{} checks to within $\qnCircuitMaxDev$
(Supplement~S5 lists them, and its Fig.~S2(c) draws this circuit). The frame checks below add two more.

\emph{Frame twirl.} The construction above is a Pauli channel by design; the frame makes \emph{every} attack one. Let
the attacker apply an arbitrary channel $\mathcal A$ to the $n$ cells of a unit. It may couple them to its own
systems, measure them and act on the outcomes, provided the same $\mathcal A$ acts on every unit and $\mathcal A$ does
not depend on the frame. In the Pauli basis $\mathcal A(\rho)=\sum_{P,Q}\chi_{PQ}\,P\rho Q^\dagger$. The verifier
applies a uniformly random Pauli $F$ at issue and $F^\dagger$ at verification, so the cells undergo
$\mathcal A_F(\rho)=F^\dagger\mathcal A(F\rho F^\dagger)F$. Because $F^\dagger PF=\chi_P(F)\,P$,
\begin{equation}
\begin{aligned}
\mathbb E_F\,\mathcal A_F(\rho)&=\sum_{P,Q}\chi_{PQ}\Bigl(4^{-n}\sum_F\chi_{P}(F)\chi_{Q}(F)\Bigr)P\rho Q^\dagger\\
&=\sum_P\chi_{PP}\,P\rho P^\dagger ,
\end{aligned}
\label{eq:twirl}
\end{equation}
since the character sum is $1$ for $P=Q$ and $0$ otherwise. The frame-averaged attack is therefore the Pauli channel
with error probabilities $\chi_{PP}$ \cite{bennett1996mixed,dankert2009exact}, and the verifier's records follow this
law because each cell's frame is drawn afresh for every unit and independently of $\mathcal A$. A coherent attack shows
what the frame does. For $\mathrm{RX}(0.7)\otimes\mathrm{RX}(0.4)$ on a two-qubit GHZ probe, exact simulation gives a
syndrome law that deviates from the Pauli-channel prediction by \qnCircuitNoFrame{} without the frame; averaged over
the 16 frames it matches (\ref{eq:twirl}) to within $\qnCircuitFrameAvg$. Two limits follow from
(\ref{eq:twirl}). A leaked frame removes the average, so the attacker can act coherently. An attack that changes from
unit to unit is twirled unit by unit, and the records then follow the mixture of the twirled channels
(Section~\ref{sec:model}).

\emph{Reading of the strength.} In this construction $\eps$ measures how well the attacker hides the conserved parity.
Without kicks, $\eps=1$; perfect hiding needs $p=1/2$ exactly. The fingerprint's support is the set of cells the
symmetric dynamics couple. A tool whose symmetry is a product of local terms has a low-weight $\astar$, which is the
regime in which we find no exponential gain (Section~\ref{sec:res-weight}).

\section{Statistics of \texorpdfstring{$\Nn$}{N90}}
\label{app:stats}
\emph{Grid.} Every trial is decoded at 40 geometrically spaced checkpoints up to its own horizon. For a group of
trials, $\Nn$ is read on a common grid of 40 geometric points from 4 units to the smallest horizon in the group. At a
grid point each trial counts as a success if it was correct at its last checkpoint at or below that point, and $\Nn$
is the first grid point at which the success fraction reaches $0.9$. A group that never reaches $0.9$ is censored and
reported as a lower bound at its horizon, never as a value.

\emph{Intervals.} We resample trials 2000 times with replacement \cite{efron1993bootstrap} and report the 5th and 95th
percentiles of the resampled $\Nn$ as a 90\% interval. Rates carry two-sided 95\% Clopper--Pearson intervals
\cite{clopper1934use}, and success curves carry 95\% Wilson bands \cite{wilson1927probable}. Tables also give an
interpolated $\Nn$, linear in $\log n$ between the two grid points around $0.9$, with its own bootstrap interval.

\emph{Fine checkpoints.} When the trials of a group have different horizons, each trial is read at a checkpoint that
can lie well below the grid point, so the grid $\Nn$ comes late. We therefore replayed the uniform-family trials (E1f)
with the same random draws and decoded them at points at most $1.03\times$ apart inside a window around $\Nn$ fixed in
advance from the stored rows. For the vault, the counts between two stored checkpoints were drawn from their exact
conditional law given the stored totals, a multivariate hypergeometric draw, so the trial's own data are unchanged.
For per-unit probes the integer sufficient statistics were truncated at each fine point. A replay check required the
decoded posterior at every stored checkpoint to equal the stored value bit for bit. Headline numbers give the fine
value first and the grid value in parentheses. Grid values from other groups with varying horizons carry
``$\pm1$ grid step''.

\emph{Crossovers.} The vault-noise crossovers (Fig.~\ref{fig:noise}) divide the vault's grid $\Nn$ by the per-unit fine
$\Nn$. Because the vault side is read on the grid, which comes late, the crossover lies on the low side.

\emph{What $\Nn$ is.} $\Nn$ is an empirical crossing of the success fraction averaged over instances drawn from the
prior. It is not a guarantee for 90\% of all instances, and some groups have 20 trials.

\section*{Acknowledgment} This work was supported in 2026 by the Information Security Core Source Technology Development Project of the Institute of Information \& Communications Technology Planning \& Evaluation (IITP) funded by the Korean government (Ministry of Science and ICT) (Project No. RS-2026-25519773, 30\%); the University Basic Research Institute Support Program (G-LAMP) of the National Research Foundation of Korea (NRF) funded by the Ministry of Education (Project No. RS-2026-25609336, 10\%); the Copyright Technology Global Talent Nurturing Program of the Korea
Creative Content Agency (KOCCA) funded by the Ministry of Science and ICT (Project No. RS-2025-02221620, 40\%); the University ICT Research Center (ITRC) program (Project No. IITP-2026-RS-2021-II211816, 10\%); and the Core Research Program of the National Research Foundation of Korea (NRF) (Project No. RS-2026-25481431, 10\%).


\onecolumn
\section*{Author biographies}

\authorbio{\includegraphics[width=\linewidth]{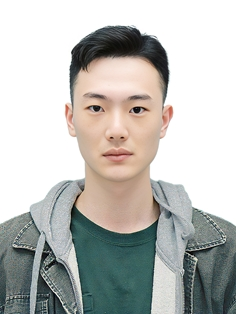}}{Jaeyun Jeong}{(Student Member, IEEE)
is currently pursuing the M.S. degree in Computer and Information Security at Sejong University,
Seoul, South Korea. His research interests include quantum information science, cybersecurity, and applied
quantum computing. His current research focuses on quantum error correction and quantum-classical hybrid artificial intelligence.}

\authorbio{\includegraphics[width=\linewidth]{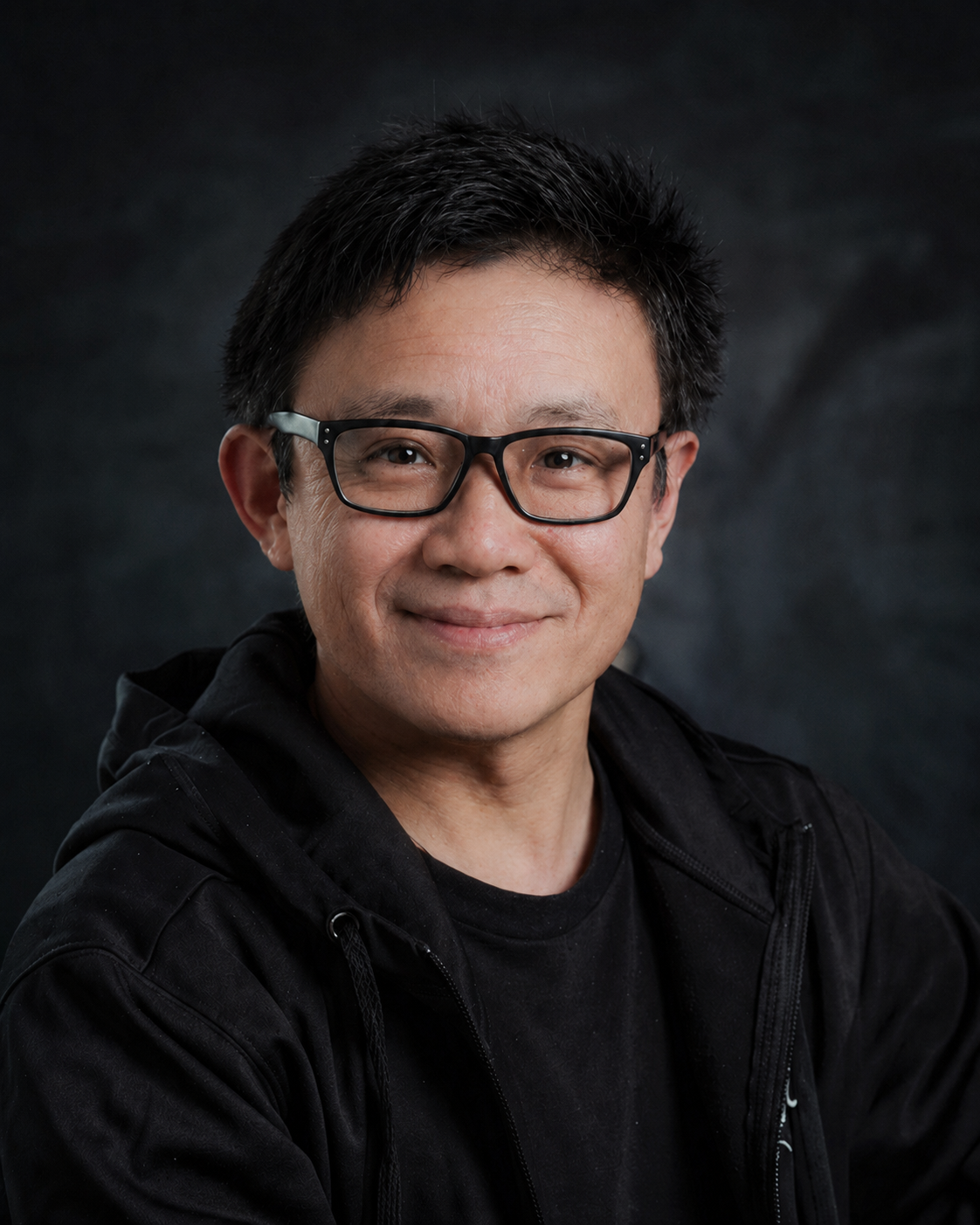}}{Zehua Wang}{(Member, IEEE) received his Ph.D. degree from the University of British Columbia (UBC), Canada, in 2016.
He is a Distinguished Professor at the China University of Mining and Technology, a Core Faculty Member of Blockchain@UBC, and an Adjunct Professor at UBC.
His research interests include blockchain, trustworthy artificial intelligence, and privacy-preserving machine learning.
He serves as an Area Editor for Computer Communications and a Youth Editorial Board Member of Blockchain: Research and Applications.
He previously served as Chief Scientist of RightMesh, which received a commendation letter from Canada’s Prime Minister in 2018.}

\authorbio{\includegraphics[width=\linewidth]{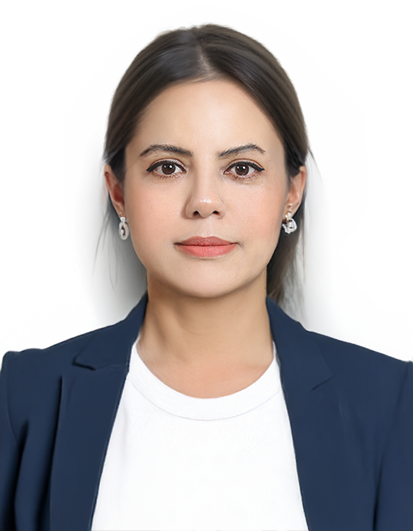}}{Lilia Tightiz}{(Member, IEEE) received her Ph.D. in computer science and engineering
from Sejong University, South Korea, in 2022. She has over 15 years of experience in
power distribution engineering. She joined Sejong University as an Assistant Professor
in 2025 after serving at Gachon University. She also serves as deputy leader of the
quantum grid team at Orientom, South Korea. Her research interests include quantum
optimization, deep reinforcement learning, quantum machine learning, smart grids,
cybersecurity, and blockchain. She serves as an Associate Editor for e-Prime–Advances
in Electrical Engineering, Electronics, and Energy, Smart Energy Systems, and
Computation, and she also serves as a reviewer for IEEE, Springer, and Elsevier
journals.}

\authorbio{\includegraphics[width=\linewidth]{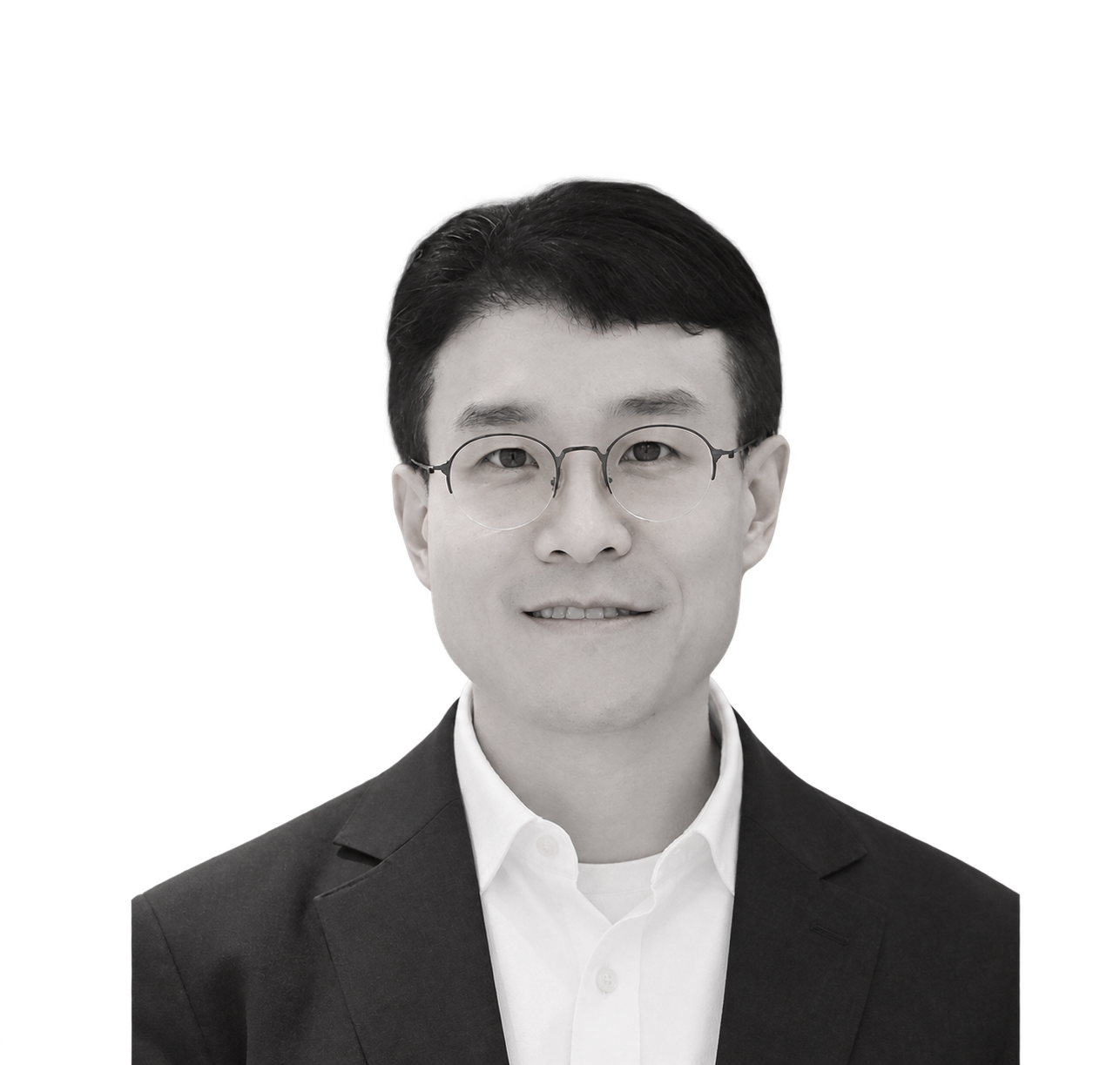}}{Ki-Woong Park}{(Member, IEEE)
received the B.S. degree in computer science from Yonsei
University and the M.S. and Ph.D. degrees in
computer science and electrical engineering from
KAIST. From 2008 to 2009, he was a Graduate
Researcher with Microsoft Research Asia and
Microsoft Research Redmond and a Researcher
with the National Security Research Institute
(NSRI), in 2012. He has a passionate interest in
designing, building, and analyzing secure systems,
especially for cloud computing systems, networked systems, and embedded
systems. He is a member of the IEEE Computer Society and ACM.}

\authorbio{\includegraphics[width=\linewidth]{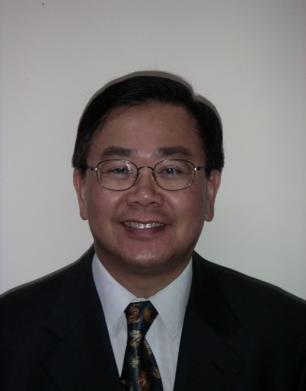}}{Victor C.M. Leung}{(Fellow, IEEE)
is an Emeritus Professor
of Electrical and Computer Engineering at the University of British Columbia,
Canada. He is also the Dean of the Artificial Intelligence Research
Institute and a Professor of Engineering at Shenzhen MSU-BIT University, China.
He has co-authored numerous refereed articles in the broad areas of wireless
networks and mobile systems, and was a Clarivate Analytics “Highly Cited Researcher”
for several years. Dr. Leung is an editor of the IEEE TGCN and several other journals.
He received the 2017 IEEE ComSoc Fred W. Ellersick Prize, 2017 IEEE Systems Journal
Best Paper Award, 2017 Canadian Award for Telecommunications Research, 2018 IEEE TCGCC
Distinguished Technical Achievement Recognition Award, 2018 IEEE CSIM Best Journal
Paper Award, and 2019 IEEE TCGCC Best Journal Paper Award, as well as several other awards.
He is a Life Fellow of IEEE, and a Fellow of the Royal Society of Canada, Canadian
Academy of Engineering, and Engineering Institute of Canada.}

\end{document}


\maketitle
\noindent The main text states the claims and its appendices give the proofs (Appendix~A), the physical realization of
the tamper model (Appendix~B) and the statistics of $\Nn$ (Appendix~C). This supplement makes the claims checkable.
Each section answers one question a reader may ask:

\begin{center}
\begin{tabular}{lp{4.6cm}p{8.2cm}}
\toprule
Section & Content & Question it answers \\
\midrule
S1 & Experiment register & Were the rules fixed before the results? \\
S2 & Claims and their evidence & What is the basis of each number? \\
S3 & Further results & What happens under other conditions? \\
S4 & Decoder and thresholds & Where do the thresholds come from? \\
S5 & Circuit checks & Does the simulator match the physics it stands for? \\
S6 & Changes to the claims & Which claims changed, and why? \\
S7 & Reproduction & Can the results be rebuilt? \\
S8 & Resource derivations & How are the engineering numbers obtained? \\
\bottomrule
\end{tabular}
\end{center}

\begin{landscape}
\section{Experiment register}
\label{sec:s1}
Every experiment, with the rule written in the laboratory notebook (available from the authors on request) before it
ran. Raw rows are
counted from the files in \path{outputs/results}. Two times are given, both in local time on
the same clock. The notebook time is the one written in the heading of the notebook section that fixed the protocol.
The run start is the earliest \texttt{START} that the run scripts logged in \path{progress.txt} for the experiment's
steps, independently of the notebook; the column after it compares the two. ``Post hoc'' marks an experiment added
after earlier results were seen; its own prediction and rule were still written before it ran. The last column lists
the claims of Table~\ref{tab:claims} that rest on the experiment. The first line of runs used older code and is
superseded; no claim rests on it.

Where a heading has a clock time (E1f, E11--E15) it precedes the logged start by at most three minutes. Most headings
give only a date, which the logs cannot order within the day. For E1 and E2 the heading carries the date 2026-11-06,
while the run log starts on 2026-11-05 at 18:10 and 22:13. For these two experiments
the notebook therefore does not document that the protocol preceded the run, and we do not claim that it does on the
strength of the record alone; their status reads ``rule written; order not documented''. The claims that rest on them
are C3, C6--C8 and C20 (E1f replays E1's trials with the same draws, so C3 and C20 rest on E1's data).

{\footnotesize
\setlength{\tabcolsep}{3pt}
\begin{longtable}{L{1.0cm}L{3.3cm}L{4.7cm}L{3.3cm}L{3.3cm}L{1.75cm}L{2.2cm}L{2.2cm}L{1.6cm}}
\caption{Experiment register.}\label{tab:register}\\
\toprule
ID & Question & Pre-registered rule & Outcome & Raw rows (files) & Seed space & Notebook time; status & Run start (log);
order & Claims (Table~\ref{tab:claims}) \\
\midrule
\endfirsthead
\toprule
ID & Question & Pre-registered rule & Outcome & Raw rows (files) & Seed space & Notebook time; status & Run start (log);
order & Claims (Table~\ref{tab:claims}) \\
\midrule
\endhead
\bottomrule
\endlastfoot
E1 & Uniform fingerprint: vault vs per-unit uniform-Lagrangian probes, $r=4$--$10$, $\varepsilon\in\{0.15,0.6\}$, $p_0\in\{0,0.005\}$ & (i) vault $N_{90}\varepsilon^2/r\le10$; (ii) per-unit $\log_2$ slope $\ge0.7$; (iii) ratio $\ge10$ at the largest $r$; withdraw the claim if $\varepsilon=0.15$ fails (ii) or (iii) & pass in all four $(\varepsilon,p_0)$ & 6{,}760 (\texttt{e1\_ml}, \texttt{e1\_vault}) & 941 & 2026-11-06; rule written; order not documented & 2026-11-05 18:10; date later than run start & C3, C20 \\
E1f & Fine checkpoints around $N_{90}$: E1 replayed with the same draws & predictions only: bit-identical replay; fine $\le$ grid in $\ge90\%$ of groups; fine CI over $\ge3$ points & first run failed the replay check (tie-breaking RNG), fixed and rerun; checks reported & 6{,}760 (\texttt{e1f\_ml}, \texttt{e1f\_vault}) & 941 (replay) + fine-point draws & 2026-11-06 20:50; pre-registered & 2026-11-06 20:52; before run & C3, C20 \\
E2 & Fingerprint weight $w\in\{1,2,3,\lceil r/2\rceil,r\}$, $r\in\{6,8,10\}$ & keep the high-weight claim if the $w=r$ ratio at $r=10$ is $\ge100$ and its slope $\ge0.7$; withdraw if $<10$ & kept with the pre-registered probes; revised by the GHZ baseline below & 4{,}240 (\texttt{e2}) & 941 & 2026-11-06; rule written; order not documented & 2026-11-05 22:13; date later than run start & C6, C7, C8 \\
E2-GHZ & GHZ-type per-unit probes at $w=r$ (same instances as E2) & written before the run: keep $2^r$ form if $\ge100$ and slope $\ge0.7$; $(3/2)^r$ form if $\ge10$; withdraw if $<10$ & high-weight claim downgraded to the $(3/2)^r$ form & 240 (\texttt{e2\_ghz}) & 941 & 2026-11-06; post hoc & --; no run step & C6 \\
E3 & Vault noise $p_v$ (depolarizing, dephasing), $r\in\{6,8,10\}$ & no decision rule; report the crossover; flag ``fragile'' if $p_v^*<0.05$ at $r=10$ & crossovers 0.2--0.3; flag not raised; analysis fix (censored points) post hoc & 7{,}200 (\texttt{e3\_dep}, \texttt{e3\_deph}, \texttt{e3b\_dep}, \texttt{e3b\_deph}) & 941 & 2026-11-06; pre-registered & 2026-11-06 03:24; date only, same day & C18 \\
E3f & Vault-noise crossovers against fine per-unit targets (analysis only) & keep the stated limits if the vault leads at 0.2 / 0.15 with fine targets & limits kept; one prediction failed (4 of 12 brackets moved down one step) & -- (analysis only) & -- & 2026-11-07 09:35; pre-registered & --; no run step & C18 \\
E5b & Adaptive GHZ heuristic with deterministic top-$k$ (replaces E5) & trigger a revision if the adaptive probe is $\ge3\times$ faster than GHZ at $r=10$ & not triggered (speed-ups 1.0--1.29 at $r=10$) & 240 (\texttt{e5b}) & 941 & 2026-11-06; pre-registered & 2026-11-06 13:33; date only, same day & C6 \\
E6 & Tier 3, vault: one switch between tools, 300 switch and 300 no-switch trials & detection $\ge0.9$, false alarms $\le0.05$, cells and onset $\ge0.9$, switch-time error rule & pass in all six $(r,\varepsilon)$ & 3{,}600 (\texttt{e6}) & 932 & 2026-11-06; pre-registered & 2026-11-06 07:33; date only, same day & C16 \\
E7 & Several hidden parities ($K\le3$), $\varepsilon$ and $K$ unknown, $r\le8$ & separation persists if the $r=8$ ratio is $\ge0.25\cdot2^r$ and its slope $\ge0.7$ & persists for all six $(K,\varepsilon)$ & 2{,}580 (\texttt{e7}) & 951 & 2026-11-06; pre-registered & 2026-11-06 07:42; date only, same day & C4 \\
E8 & Tier 3, per-unit detectors: blind and registry & blind detector fails at the vault budget; registry detector reported as a boundary & blind 0.00 in all settings; registry detects 1.00 (boundary) & 2{,}400 (\texttt{e8}) & 932 & 2026-11-06; pre-registered & 2026-11-06 07:39; date only, same day & C17 \\
P/local & Rerun of P and local tampering on the revised code & local: ``within $2\times$ means no advantage''; P: prediction $2/\varepsilon^2$ & local rule not met (constant factor, flat in $r$); P as predicted & 760 (\texttt{pl\_paired}, \texttt{pl\_local}) & 941 & 2026-11-06; post hoc & 2026-11-06 11:00; date only, same day & C8 \\
E9 & In-seal references vs vault under oblivious, mirror and extend (32 cells) & (a) in-seal $\le2\times$ vault; (b) in-seal and P blind ($\le0.05$), vault $\le1.5\times$; (c) in-seal blind, vault given $R$ $\le1.5\times$, vault without $R$ $\le4\times$ & (a)--(c) pass & 7{,}000 (\texttt{e9}) & 941 + tag 9/10 & 2026-11-06; pre-registered & 2026-11-06 16:38; date only, same day & C10, C12, C15 \\
E10 & Coverage bound for non-adaptive stabilizer probes (enumeration, integer program, Fano) & GHZ optimal where enumeration attains $2^{r-1}+[r\ \mathrm{even}]$; separation if bound/vault $\ge10$ at $r=10$ & GHZ optimal for $r\le9$; bound $\ge10\times$ (restated as proved ratios post hoc) & -- (analysis only) & deterministic & 2026-11-06; pre-registered & 2026-11-06 18:47; date only, same day & C2, C5, C7, C11 \\
E11 & Spray attacker: vault, in-seal, in-seal oracle (32 cells) & (a) vault unaffected $\le1.5\times$; (b) in-seal blind; (c1) oracle blind for $a^*\notin C$ & all rules pass & 4{,}800 (\texttt{e11}) & 941 + tag 11/12 & 2026-11-07 09:35; pre-registered & 2026-11-07 09:35; before run & C13 \\
E12 & Reference-free product probe vs spray & (a) not blind at any $(r,\varepsilon)$; (b) ratio to the vault in $[0.67,1.5]$ & (a) pass; (b) failed at $r=4$, $\varepsilon=0.6$ & 1{,}600 (\texttt{e12}) & 941 + tag 13/14 & 2026-11-07 12:13; pre-registered & 2026-11-07 12:13; before run & C14 \\
E13 & Self-referencing P + sign; its sample bound under extend & P + sign learns $(a^*,s)$; sign phase within the stated factor; ratio to the vault near $2/\varepsilon^2$ & all rules pass & 560 (\texttt{e13}) & 941 + tag 15 & 2026-11-08 00:58; pre-registered & 2026-11-08 00:59; before run & C9, C11 \\
E14 & Seal noise $p_0\le0.1$ (vault noise fixed at 0.005): vault $r=6$--$10$, per-unit $r=6,8$ & P1: $N_{90}(p_0)/N_{90}(0.005)$ CI overlaps $[A_{3r/4},A_r]$; P2: per-unit/vault ratio within $\times1.5$; P3: control matches E1f; D1--D3 fix the wording & P1 35/36 (vault $r=6$, $\varepsilon=0.15$, $p_0=0.01$ below the band); P2 12/12; P3 10/10 & 6{,}800 (\texttt{e14\_ml}, \texttt{e14\_vault}) & E1 instances + tag 961 & 2026-11-10 14:54; pre-registered & 2026-11-10 14:54; before run & C21 \\
E15 & Tier 3 under transit reordering: E6's vault records with a fraction $f\le1$ of the units permuted & E6's rule for each $f$; control: $f=0$ equals E6; $f^*$ = largest $f$ that keeps the rule in all six settings & control equal; $f^*=0.05$; $f=0.2$: 4 of 6 fail, false switches at $\varepsilon=0.15$; $f\ge0.5$: all fail & 18{,}000 (\texttt{e15}) & 932 + tag 935 & 2026-11-11 14:05; post hoc & 2026-11-11 14:06; before run & C22 \\

\end{longtable}}
\end{landscape}

\section{Claims and their evidence}
\label{sec:s2}
Table~\ref{tab:claims} lists every headline claim with its verifier class, the kind of evidence and the source in the
repository. ``Proved'' means a lower bound divided by an upper bound, or a reduction; ``simulated'' means an $\Nn$
ratio observed in simulation; ``bound/simulated'' locates a bound and proves nothing.

{\small
\setlength{\tabcolsep}{3pt}
\begin{longtable}{L{0.6cm}L{3.5cm}L{2.3cm}L{3.0cm}L{3.0cm}L{3.3cm}}
\caption{Claims, evidence and sources ($r=10$, $p_0=0.005$ unless stated). The experiments behind each claim are in
the last column of Table~\ref{tab:register} (Section~\ref{sec:s1}).}\label{tab:claims}\\
\toprule
\# & Claim & Class & Evidence & Value & Source \\
\midrule
\endfirsthead
\toprule
\# & Claim & Class & Evidence & Value & Source \\
\midrule
\endhead
C1 & Uniform fingerprint: exponential separation & per-unit, no untouched reference & proved by reduction
($\eps\le1/3$, one parity); asymptotic & $\Omega(2^r/\eps^2)$; explicit: $N\ge$\qnChenLbRten{} at $r=10$,
$\eps=0.15$ (vault upper bound \qnChenVaultUbRten), separation from $r=\qnChenRstar$ & Proposition~1 and its explicit form (main text, Section~V-A);
Chen \emph{et al.}, SM S3, Eqs.~(S67), (S69) \\
C2 & Same, non-adaptive stabilizer probes & per-unit & proved ratio & $\ge$\qnProvedUniform & \texttt{scope.json},
\texttt{t11\_bound} \\
C3 & Same, implemented strategy & per-unit & simulated (fine; grid) & \qnEoneRatioFine{} (\qnEoneRatioGrid) &
E1, E1f; \texttt{t1\_uniform} \\
C4 & Several parities, $\eps$, $K\le3$ unknown & per-unit & simulated, $r=8$ & \qnEsevenRatio & E7; Fig.~\ref{fig:multi} \\
C5 & Full weight $w=r$ & per-unit, non-adaptive stabilizer & proved ratio (theorem $c^*$) & $\ge$\qnProvedWr{}
($\ge$\qnProvedWrThm) & E10; \texttt{t11\_bound} \\
C6 & Full weight $w=r$ & per-unit stabilizer, incl.\ adaptive & simulated & \qnWrMeas & E2, E5b; \texttt{t2\_weight} \\
C7 & Mid weight $w=\lceil r/2\rceil$ & per-unit, non-adaptive stabilizer & proved ratio; simulated & $\ge$\qnProvedWhalf;
\qnWhalfMeas & E10, E2 \\
C8 & Low weight $w\le3$, local tamper & per-unit & simulated & $\le3^w$; local \qnLocalRatio & E2; \texttt{t8\_local} \\
C9 & Implemented P + sign & self-referencing & simulated (fine) & \qnPsignOverVaultFineLow{} ($\eps=0.15$),
\qnPsignOverVaultFineHigh{} ($\eps=0.6$) & E13; \texttt{t14\_self\_ref} \\
C10 & Oblivious attacker, untouched cells & in-seal reference & simulated & \qnInsealObl$\times$ the vault & E9;
\texttt{t9\_references} \\
C11 & Extend & per-unit vs vault, $R$ unknown & proved ratio & $\ge$\qnProvedExtendPerUnit & E13 analysis;
\texttt{t13\_extend} \\
C12 & Extend, vault given $R$ & protected reference & simulated & \qnExtendGivenR$\times$ oblivious & E9 \\
C13 & Spray, in-seal oracle, $\astar\notin C$ & in-seal reference & simulated & \qnOracleNotInC{} at the horizon & E11;
\texttt{t12\_spray} \\
C14 & Spray, product probe & per-unit (no reference) & simulated & \qnProductRatio$\times$ the vault & E12 \\
C15 & Mirror & in-seal; P & simulated & \qnMirrorInseal; at most \qnMirrorPaired & E9 \\
C16 & Tier 3, vault & protected reference & simulated & detection $\ge$\qnTierThreeVaultDetect, \qnTierThreeVaultFA{}
false alarms & E6; \texttt{t5\_tier3} \\
C17 & Tier 3, blind / registry & per-unit & simulated & \qnTierThreeBlind{} / \qnTierThreeRegistry & E8 \\
C18 & Vault noise limit & protected reference & simulated, finite size & leads up to $p_v=0.2$ (dep.), 0.15 (deph.) &
E3, E3f; \texttt{t4\_vault\_noise} \\
C19 & Circuits match the sampler & -- & exact simulation & \qnCircuitChecks{} checks, max.\ dev.\ $\qnCircuitMaxDev$ &
S5 \\
C20 & Attribution threshold and vault inventory & protected reference, per-unit & simulated (fine) & vault \qnVaultNninetyRtenLow{} units, \qnVaultQubitsRtenLow{} vault qubits & E1, E1f; main-text Table~VI \\
C21 & Seal requirement curve: cost at seal noise $p_0$ & protected reference, per-unit & simulated; attenuation law
pre-registered & vault $\times$\qnSealRhoFive{} at $p_0=0.05$, $\times$\qnSealRhoTen{} at $0.1$ ($r=10$); per-unit / vault
ratio $\times$\qnSealRatioChange; law \qnSealPOne{} groups & E14; \texttt{t15\_seal}; main-text Fig.~8 \\
C22 & Tier 3 under transit reordering & protected reference & simulated, post hoc & E6's rule kept with up to
\qnReorderFstar{} of the units reordered; false switches from 20\% (all from negative-sign fingerprints) & E15;
\texttt{t16\_reorder} \\
\bottomrule
\end{longtable}}

\section{Further results}
\label{sec:s3}
\begin{figure}[h]
\centering
\includegraphics[width=0.75\textwidth]{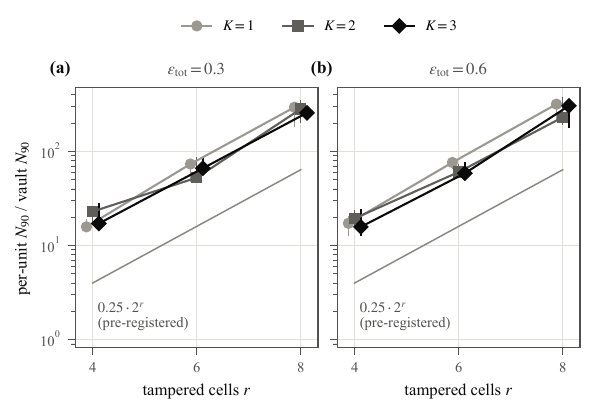}
\caption{Several hidden parities, with $\eps$ and $K$ unknown to both laboratories (E7). $\Nn$ ratio of the per-unit
strategy to the vault, with intervals from the CI endpoints. The gray line is the pre-registered persistence threshold
$0.25\cdot2^r$.}
\label{fig:multi}
\end{figure}

\emph{Several parities (E7).} Fig.~\ref{fig:multi} shows the per-unit/vault ratio when the channel carries $K\le3$
hidden parities with $\eps$ and $K$ unknown to both laboratories. Because a mixture of Pauli channels averages their
eigenvalues, this is also the channel of an attacker who picks one of $K$ tools at random for each unit (main text,
Section~III). The vault identified all $K$ parities in every setting, and its cost grows roughly as $K^2$; the ratio
persists at \qnEsevenRatio{} at $r=8$.

\emph{Tier 3 under transit reordering (E15, post hoc).} The protocol and decision rules were written before the run.
E6's vault records (300 switch and 300 no-switch trials for each $r\in\{6,8,10\}$ and
$\eps\in\{0.15,0.6\}$) were regenerated from their seeds, and a fraction $f\in\{0,0.05,0.2,0.5,1\}$ of each record's
units, at positions drawn uniformly, was permuted among those positions before the report read the record; $f=1$ is a
full permutation. At $f=0$ all \qnReorderControl{} rows equal E6's stored rows. E6's rule (detection $\ge0.9$,
false-alarm upper bound $\le0.05$, cells exact $\ge0.9$) holds in all six settings up to $f=0.05$ (detection
\qnReorderDetectFive); at $f=0.2$ \qnReorderFailTwenty{} settings fail (detection \qnReorderDetectTwenty, up to
\qnReorderFATwenty{} false switches), and at $f=1$ detection is \qnReorderDetectFull{} with \qnReorderFAFull{} false
switches out of 300. The cell error rates do not depend on the order, so tier~1 localized the cells in every trial.
A diagnosis written after the results splits the trials by the sign of the true fingerprint: all \qnReorderFATotal{}
false switches came from negative-sign fingerprints. Untampered units mixed in after the onset show the identity, whose
parities are all $+1$; they lift every parity by their share, pull a negative fingerprint toward zero, and let each
window decode a spurious positive parity. Table \texttt{t16\_reorder} lists every group.

\emph{Tables in the repository.} The following tables in \texttt{outputs/paper\_tables} (CSV and \texttt{tables.md}) are
built from the raw rows and hold the results not plotted in the main text:
\texttt{t1\_uniform} (E1 with $p_0=0$ and $0.005$, grid, interpolated and fine $\Nn$ with intervals),
\texttt{t2\_weight} (all weight families), \texttt{t3\_adaptive} (E5b), \texttt{t4\_vault\_noise} (E3, E3f),
\texttt{t5\_tier3} (E6, E8), \texttt{t6\_multi} (E7), \texttt{t7\_paired} (P alone), \texttt{t8\_local} (local
tampering), \texttt{t9\_references} (E9), \texttt{t10\_coverage} (E10 coverage), \texttt{t11\_bound} (Fano bounds and
proved ratios), \texttt{t12\_spray} (E11, E12), \texttt{t13\_extend} (extend bounds), \texttt{t14\_self\_ref} (E13),
\texttt{t15\_seal} (E14: every group's $\Nn$, cost factor, predicted band and the P1 result) and
\texttt{t16\_reorder} (E15).
The figure script reproduces the $p_0=0$ variants of main-text Fig.~3 (\texttt{fig2(..., p0=0.0)}) and success curves at other
$r$.

\section{Decoder and thresholds}
\label{sec:s4}
\emph{Status logic of \texttt{qmf\_report}.} The report returns \texttt{global\_anomaly} when the median cell rate
exceeds $\tau(b_0)$; \texttt{clean} when no cell exceeds $\tau(\tilde\rho)$; \texttt{copied\_errors} when two
localized cells pass the copy test; \texttt{not\_decoded} when more than 12 cells are localized; and
\texttt{localized} otherwise. Every report carries a \texttt{tier3} field, \texttt{not\_tested} for the early returns.

\emph{Localization, $\tau(x)=x+5\sqrt{\max(x,1/N)/N}+5/N$.} Five binomial standard deviations keep the chance that a
benign cell is flagged below about $3\times10^{-7}$ per cell under a normal approximation; the floor $1/N$ and the term
$5/N$ guard small $N$ and rates near zero.

\emph{Onset.} The maximum-likelihood single change point of the indicator ``any error on $\hat R$'', with Bernoulli
rates fitted on both sides.

\emph{Copies.} Two cells are copies if their codes agree on at least $0.8$ of the units where either is not the
identity, counted over at least 20 such units. Independent scrambled cells agree on such units with probability about
$0.2$; exact copies agree except for benign noise. Counting only units with an error keeps two rarely hit cells from
looking like copies.

\emph{Tier 2.} The fingerprint is significant if
$|\hat\lambda_{\hat a}|>\sqrt{2\ln(2\cdot4^r/\delta)/N_{\mathrm{post}}}$ with $\delta=0.01$, where $N_{\mathrm{post}}$
is the number of units from the onset on; the window fingerprints of tier~3 use the same test with $W$ in place of
$N_{\mathrm{post}}$. The threshold is Hoeffding's inequality with a Bonferroni correction over the $4^r$ candidates and
both signs. It bounds the chance of a spurious significant fingerprint by $\delta$ for a fixed region and onset; both
are chosen from the same records, so the level is nominal.

\emph{Tier 3.} The switch statistic $G_k=\sum_{u\ge k}g_u$ uses the log-likelihood ratio of the last-window fingerprint
against the first. A switch is declared if $\max_kG_k>2\ln N_{\mathrm{post}}$, a penalty of the order of the
log-likelihood gain of a spurious change point. Registry links require a $z$-score above 4. The experiments use the
window $W=10r/\eps^2$; the code's default is a third of the record.

\emph{What the thresholds guarantee.} Only the tier-2 threshold is derived; the others are heuristics whose behaviour
was simulated or tested, not derived.

\begin{center}
\small
\begin{tabular}{L{4.0cm}L{4.4cm}L{7.0cm}}
\toprule
Test & Threshold & Basis \\
\midrule
Global anomaly, localization & $\tau(x)$ & Heuristic: five standard deviations under a normal approximation;
cells and onset simulated in E6 \\
Copies & agreement $\ge0.8$ on $\ge20$ units & Heuristic; unit tests \\
Tier-2 significance & $\sqrt{2\ln(2\cdot4^r/\delta)/N_{\mathrm{post}}}$ & Hoeffding and Bonferroni, $\delta=0.01$, for a
fixed region and onset \\
Tier-3 switch & $2\ln N_{\mathrm{post}}$ & Heuristic penalty; false alarms simulated in E6 \\
Registry link & $z>4$ & Heuristic; unit tests \\
\bottomrule
\end{tabular}
\end{center}

\noindent Each test has its own decision rule, but only the tier-2 threshold admits a theoretical error bound, and that
bound is conditional on the region and onset. The whole report has no established family-wise error guarantee, and the
tier-3 false-alarm count of \qnTierThreeVaultFA{} per setting is a simulation result for the E6 scenario, not a guarantee.

\section{Circuit checks}
\label{sec:s5}
\begin{figure}[p]
\centering
\begin{minipage}[b]{0.42\textwidth}\centering
\includegraphics[height=4.6cm]{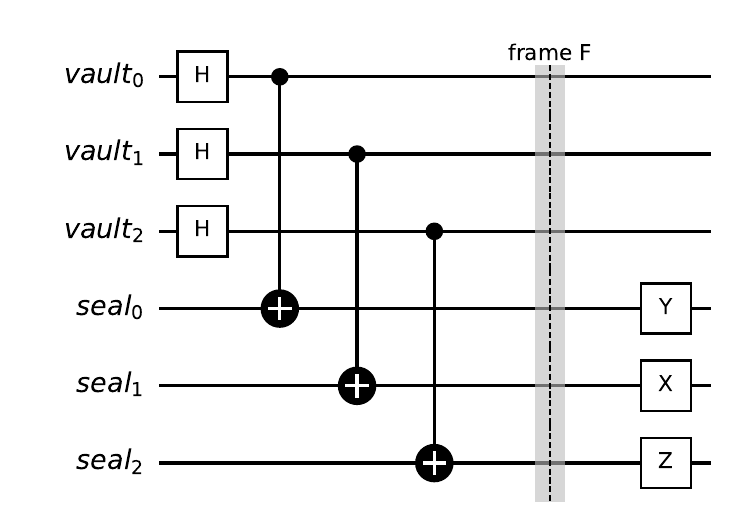}\\[2pt](a)
\end{minipage}\hfill
\begin{minipage}[b]{0.56\textwidth}\centering
\includegraphics[height=4.6cm]{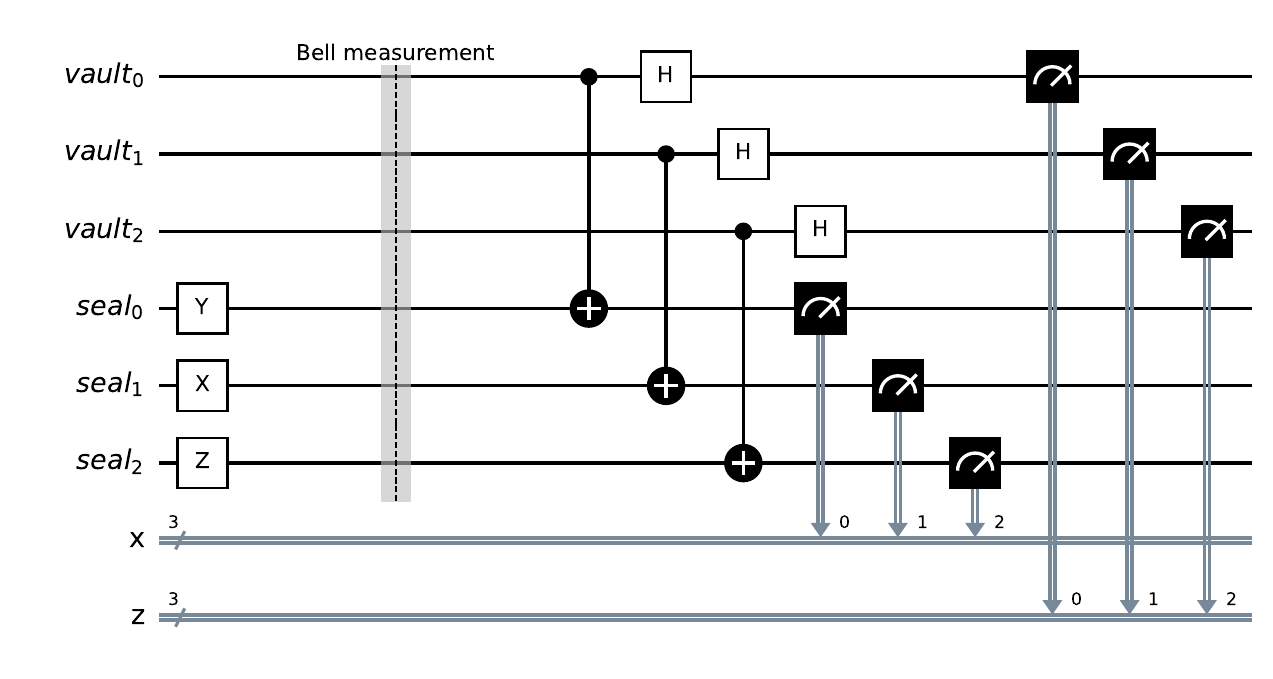}\\[2pt](b)
\end{minipage}

\vspace{10pt}
\includegraphics[height=6.4cm]{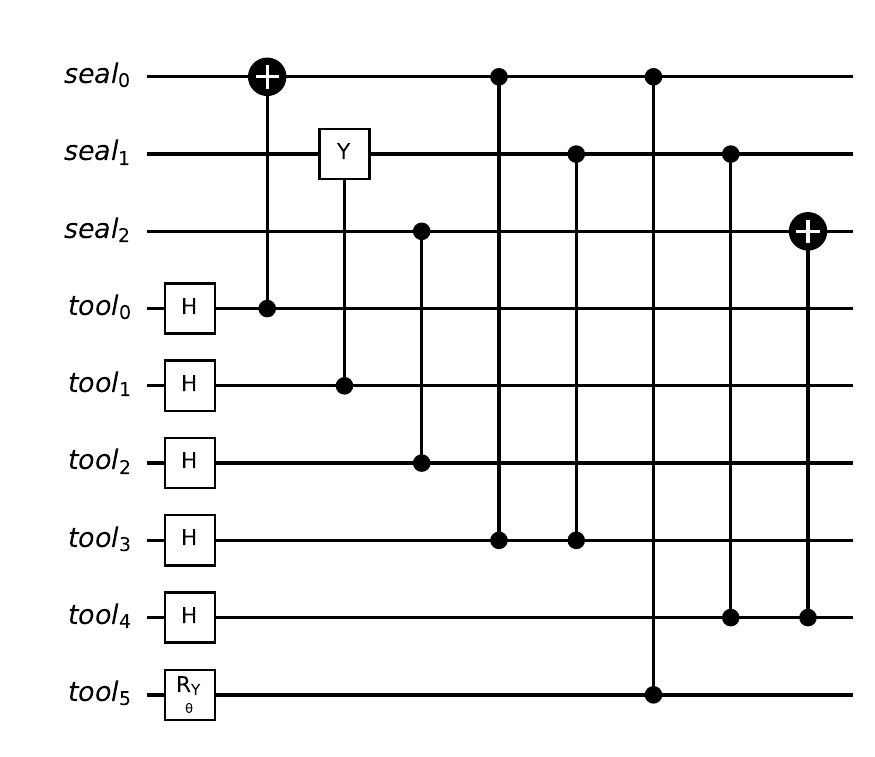}\\[2pt](c)

\vspace{10pt}
\includegraphics[width=0.95\textwidth]{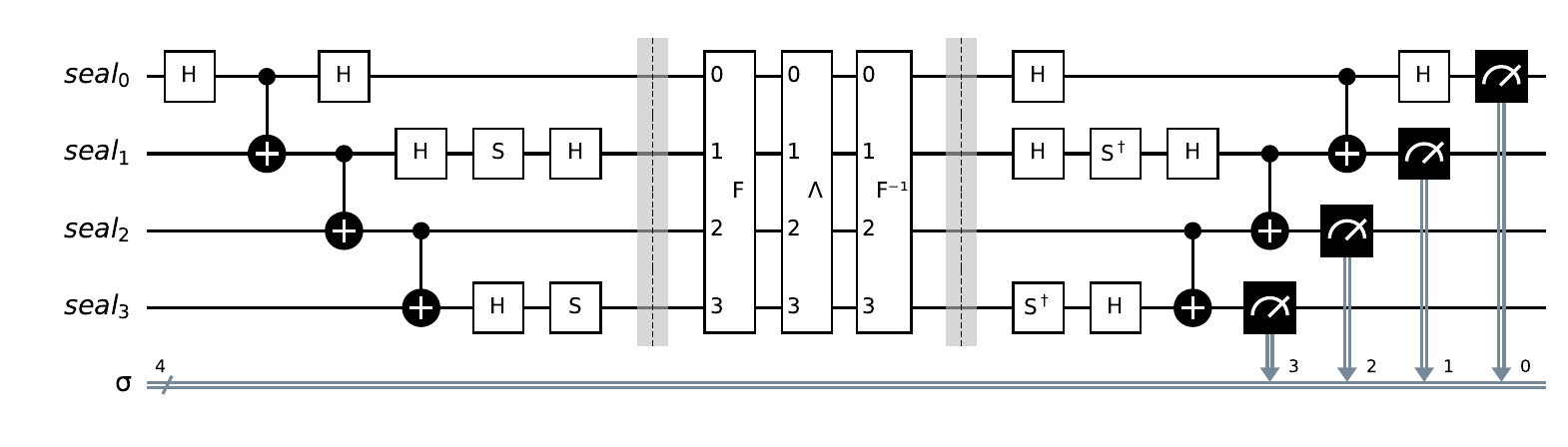}\\[2pt](d)
\caption{The circuits of the checks for $r=3$ ((a)--(c)) and $r=4$ ((d)).
(a)~Issue: Bell pairs between vault and seal cells, then the secret frame $F$ on the seal half (here
$F=Y\otimes X\otimes Z$).
(b)~Verification: the frame is undone, and each seal cell is Bell-measured with its vault partner; the bits $x$ and
$z$ give the Pauli error $E_u$.
(c)~The tamper tool of Appendix~B for $\astar=X\otimes Y\otimes Z$: tool qubits 0--4, prepared in $|+\rangle$, control
the generators $X_0$, $Y_1$, $Z_2$, $Z_0Z_1$ and $Z_1X_2$ of the centralizer of $\astar$; tool qubit 5, prepared by
$R_Y(\theta)$ with $\sin^2(\theta/2)=(1-s\eps)/2$, controls $Q=Z_0$. Discarding the tool qubits leaves the channel of
Eq.~(1) of the main text on the seal cells.
(d)~A per-unit GHZ-type probe on four cells with local Clifford frames; $F$, $\Lambda$ and $F^{-1}=F^\dagger$ are the
frame, the attack and the frame's inverse, and the measured bits are the syndromes $\sigma$.}
\label{fig:circuits}
\end{figure}

Exact Qiskit simulation (statevector or density matrix) of the issue, tamper, probe and verification blocks, compared
with the sampler's closed-form laws. Pass bar $10^{-10}$; the coherent-attack check without the frame is expected to
deviate. Source: \path{outputs/paper_figure/circuits/circuit_check.json}; tests
\path{tests/test_circuits.py}. Fig.~\ref{fig:circuits} draws the blocks with the same code. Panels (a)--(c) equal, as unitaries,
the blocks the checks simulate; panel (d) uses the preparation of the GHZ check and draws the frame,
the attack and the inverse frame as boxes.

{\small
\setlength{\tabcolsep}{4pt}
\begin{longtable}{rL{4.0cm}L{6.6cm}lll}
\caption{The \qnCircuitChecks{} circuit-versus-law checks.}\label{tab:circuits}\\
\toprule
\# & Check & Parameters & Expected & Max.\ deviation & OK \\
\midrule
\endfirsthead
\toprule
\# & Check & Parameters & Expected & Max.\ deviation & OK \\
\midrule
\endhead
\bottomrule
\endlastfoot
1 & vault, noiseless & r=1, a=2, s=1, eps=0.15, frame=0 & 0 & $2.2\times10^{-16}$ & yes \\
2 & vault, noiseless & r=1, a=1, s=1, eps=0.6, frame=2 & 0 & $2.8\times10^{-16}$ & yes \\
3 & vault, noiseless & r=1, a=1, s=-1, eps=0.15, frame=3 & 0 & $2.2\times10^{-16}$ & yes \\
4 & vault, noiseless & r=1, a=1, s=1, eps=0.6, frame=3 & 0 & $2.8\times10^{-16}$ & yes \\
5 & vault, noiseless & r=2, a=5, s=-1, eps=0.15, frame=8 & 0 & $8.3\times10^{-17}$ & yes \\
6 & vault, noiseless & r=2, a=8, s=1, eps=0.6, frame=12 & 0 & $1.4\times10^{-16}$ & yes \\
7 & vault, noiseless & r=2, a=14, s=1, eps=0.6, frame=6 & 0 & $1.4\times10^{-16}$ & yes \\
8 & vault, noiseless & r=2, a=7, s=-1, eps=0.6, frame=8 & 0 & $1.4\times10^{-16}$ & yes \\
9 & vault, noiseless & r=3, a=50, s=1, eps=0.6, frame=59 & 0 & $4.5\times10^{-17}$ & yes \\
10 & vault, noiseless & r=3, a=53, s=-1, eps=0.15, frame=35 & 0 & $3.5\times10^{-17}$ & yes \\
11 & vault, noiseless & r=3, a=21, s=-1, eps=0.15, frame=40 & 0 & $3.5\times10^{-17}$ & yes \\
12 & vault, noiseless & r=3, a=3, s=1, eps=0.6, frame=23 & 0 & $4.5\times10^{-17}$ & yes \\
13 & vault, benign noise & r=1, a=1, s=1, eps=0.6, frame=0, p0=0.05, pv=0.2, vkind=dep & 0 & $3.3\times10^{-16}$ & yes \\
14 & vault, benign noise & r=2, a=11, s=1, eps=0.6, frame=15, p0=0.05, pv=0.2, vkind=dep & 0 & $1.1\times10^{-16}$ & yes \\
15 & vault, benign noise & r=1, a=3, s=1, eps=0.6, frame=2, p0=0.05, pv=0.2, vkind=deph & 0 & $2.8\times10^{-16}$ & yes \\
16 & vault, benign noise & r=2, a=13, s=-1, eps=0.6, frame=9, p0=0.05, pv=0.2, vkind=deph & 0 & $1.4\times10^{-16}$ & yes \\
17 & GHZ probe, noiseless & r=2, a=8, s=1, eps=0.6, frame=0, Mi=[0, 1] & 0 & $3.3\times10^{-16}$ & yes \\
18 & GHZ probe, noiseless & r=2, a=2, s=-1, eps=0.6, frame=15, Mi=[4, 1] & 0 & $3.9\times10^{-16}$ & yes \\
19 & GHZ probe, noiseless & r=2, a=11, s=-1, eps=0.6, frame=2, Mi=[0, 1] & 0 & $3.1\times10^{-16}$ & yes \\
20 & GHZ probe, noiseless & r=3, a=3, s=1, eps=0.6, frame=6, Mi=[4, 5, 1] & 0 & $2.4\times10^{-16}$ & yes \\
21 & GHZ probe, noiseless & r=3, a=42, s=1, eps=0.6, frame=63, Mi=[3, 2, 3] & 0 & $3.3\times10^{-16}$ & yes \\
22 & GHZ probe, noiseless & r=3, a=43, s=-1, eps=0.6, frame=61, Mi=[3, 5, 1] & 0 & $2.9\times10^{-16}$ & yes \\
23 & GHZ probe, noiseless & r=4, a=220, s=-1, eps=0.6, frame=187, Mi=[4, 5, 5, 0] & 0 & $1.5\times10^{-16}$ & yes \\
24 & GHZ probe, noiseless & r=4, a=59, s=-1, eps=0.6, frame=110, Mi=[4, 2, 0, 3] & 0 & $1.5\times10^{-16}$ & yes \\
25 & GHZ probe, noiseless & r=4, a=192, s=1, eps=0.6, frame=34, Mi=[4, 1, 3, 5] & 0 & $2.1\times10^{-16}$ & yes \\
26 & GHZ probe, seal noise & r=2, a=10, s=1, eps=0.6, frame=0, p0=0.05, Mi=[3, 5] & 0 & $5.3\times10^{-16}$ & yes \\
27 & GHZ probe, seal noise & r=3, a=25, s=-1, eps=0.6, frame=5, p0=0.05, Mi=[0, 3, 2] & 0 & $2.1\times10^{-16}$ & yes \\
28 & coherent attack, no frame & r=2, attack=RX(0.7) x RX(0.4) & $>$ 1e-3 & $1.3\times10^{-1}$ & yes \\
29 & coherent attack, frame avg & r=2, attack=RX(0.7) x RX(0.4) & 0 & $3.3\times10^{-16}$ & yes \\

\end{longtable}}

\section{Changes to the claims}
\label{sec:s6}
These claims changed during the analysis.
\begin{enumerate}
\item \emph{First line superseded.} The first line of runs had older seeds, tie-breaking on the data RNG and a
batching bug that favoured the per-unit side. Every experiment behind a claim was rerun on the corrected code.
\item \emph{Full weight (E2 and the GHZ baseline).} The pre-registered probes supported a $2^r$-type separation at
$w=r$. A GHZ-type baseline, added post hoc with its rule written before it ran, covers a weight-$r$ Pauli with
probability $\Theta((2/3)^r)$; the claim was downgraded to the $(3/2)^r$ form.
\item \emph{References and attacks (E9--E12).} Untouched seal cells turned out to be a reference as good as the vault
against an oblivious attacker, so Proposition~1 was scoped to verifiers without one. An initial expectation was that
spray would push in-seal references into the exponential regime; the pre-registered analysis corrected this (blind exactly
for $\astar\notin C$, vault-like otherwise), and E12 showed that copies inform reference-free product probes.
\item \emph{Self-referencing (E13).} P alone carries no sign; P + sign learns $(\astar,s)$, so the claim against
self-referencing verifiers became a simulated factor of about $2/\eps^2$ for the implemented strategy, with no
class-wide bound.
\item \emph{Proved ratios.} Ratios of a lower bound to a simulated cost were replaced by ratios of a lower bound to an
upper bound; proved lower bounds are rounded down.
\item \emph{Range of Proposition~1.} Reading Chen \emph{et al.}'s supplement showed that their condition
$\eps_{\mathrm C}\le1/6$ is on the estimation precision, and the hard instance's eigenvalue is $\eps_0=2\eps_{\mathrm
C}\le1/3$. Our strength is that eigenvalue, so the proved range is $\eps\le1/3$, not $\eps\le1/6$. No decision or
simulation result changed: $\eps=0.15$ is inside either range and $\eps=0.6$ outside both.
\item \emph{Explicit constant of Proposition~1.} With the constant of Chen \emph{et al.}'s Eqs.~(S67)
and (S69), the reduction gives $N\ge$\qnChenLbRten{} at $r=10$ and $\eps=0.15$, below the vault's upper bound of
\qnChenVaultUbRten. Proposition~1 is therefore stated as asymptotic: it separates every per-unit verifier from the vault
only from $r=\qnChenRstar$, and the proved ratios at $r\le10$ rest on the Fano bound for non-adaptive stabilizer probes.
\item \emph{Fingerprint erasure.} An attacker with coherent access can erase the fingerprint exactly with a uniformly
random Pauli layer on the tampered cells, drawn afresh for every unit, at the cost of single-qubit gates and random
bits. The claim that erasing the fingerprint would cost the attacker many tools was withdrawn: forensics concerns
attackers who do not erase their fingerprint, and a biased layer leaves a residual $\eps\beta$ at about $\beta^{-2}$
times the attribution cost.
\item \emph{Seal lifetime (E14).} A conclusion that the seal, not the vault, limits the lifetime rested on the single
seal noise $p_0=0.005$ that every earlier run assumed. E14, pre-registered, simulated the cost of
larger seal noise: it follows the seal attenuation law (\qnSealPOne{} groups), leaves the per-unit / vault ratio
within the pre-registered $\times1.5$ (\qnSealPTwo), and gives a curve: $p_0=0.05$ lengthens the seal lifetime
\qnSealLifeGainFive-fold at \qnSealRhoFive{} times the vault's cost at $r=10$. The conclusion is now stated as that
trade-off.
\item \emph{Attribution scope.} Attribution is stated up to fingerprint equivalence: tools that leave the same
fingerprint, and an attacker who reproduces another tool's fingerprint, are not told apart, and a registry link means
the same fingerprint on the same cells, not the same device.
\item \emph{Arrival order (E15, post hoc).} Tier~1's onset and tier~3's switch are change points in
arrival order. E15 found that tier~3 keeps E6's rule with up to \qnReorderFstar{} of the units reordered and raises
false switches beyond, all from negative-sign fingerprints, because untampered units mixed in after the onset lift
every parity. The full report is now stated for tampered units that follow the onset contiguously (main text,
Section~IX-B).
\end{enumerate}

\section{Reproduction}
\label{sec:s7}
\emph{Environments.} Results, tables and figures: Python 3.13.9, numpy 2.3.5, scipy 1.16.3, matplotlib 3.10.6, pytest
8.4.2. Circuits: conda environment with Python 3.12.0, numpy 2.5.0, Qiskit 2.4.2, qiskit-aer 0.17.2. Both
environments pass the repository check.

\emph{Commands} (from the repository root):
\begin{verbatim}
pip install -r requirements.txt
pip install -e .                                 # the qmf package (src/)
python -m pytest -q -p no:cacheprovider tests    # unit tests
python -m qmf.reproduce                          # full check (~15 min)
python -m qmf.paper.figures                      # figures (~3 min)
python -m qmf.paper.numbers                      # outputs/numbers/qmf_numbers.tex
conda run -n quantum python -m qmf.circuits \
    check outputs/paper_figure/circuits/circuit_check.json
\end{verbatim}

The plotted numbers of every figure are in \path{outputs/paper_figure/source_data/} of the repository.
The full check regenerates sample rows from their seeds, every summary, the tables, the figure data and the headline
numbers, and compares them with the stored files; with Qiskit installed it also reruns the
circuit checks.

\emph{Rerunning the experiments.} Each experiment's grid is a config file, \path{configs/*.json}, run by
\texttt{python -m qmf.run configs/e14.json} (for example). The runs used 6 workers on 8 cores with single-threaded BLAS.
The longest is E1--E3 (about 11 hours); E14 takes about 7.5 hours (most of it the vault at $r=10$, where every
checkpoint decodes all $4^{10}$ candidates), E9 about 2.2 hours and E13 about 15 minutes. The repository's README lists
every config with its running time.

{\sloppy %
\emph{Integrity.} The file \path{SHA256SUMS} at the repository root holds the SHA-256 of every file; check it with
\texttt{sha256sum -c SHA256SUMS}. Deposit: Zenodo, release v1.0.0,
doi:~\href{https://doi.org/10.5281/zenodo.24148790}{10.5281/zenodo.24148790} (published 2026-11-12). Repository:
\url{https://github.com/Jeong-Jaeyun/Quantum-Memory-Forensics-for-Tampered-Quantum-Seals-Resource-Dependent-Sample-Gains}.\par}

\section{Resource derivations}
\label{sec:s8}
\emph{Attribution threshold and inventory.} With all $N$ modules of a consignment outstanding together, the vault
holds $nN$ qubits, each for its module's round-trip time $T_u$, so the qubit-time is $n\sum_uT_u\le nNT$. Attribution
with probability $0.9$ needs $N_{\mathrm{tamp}}\ge\Nn$ tampered modules; if the attacker touches a fraction $f$ of the
consignment, $N\ge\Nn/f$, for a decoder that uses only the tampered modules. The full report assumes that every unit
after the onset is tampered (main text, Section~IX-B). With continuous issue at rate $\lambda$, the number of modules in transit is $\lambda T$
(Little's law), the inventory is $n\lambda T$, and attribution waits until $\Nn$ tampered modules have arrived.

\emph{Storage times.} With exponential decay, a dephasing memory flips with probability
$p(T)=\tfrac12(1-e^{-T/T_2})$ and a depolarizing one has $p(T)=\tfrac34(1-e^{-T/\tau})$. The vault limits
$p_v\le0.15$ (dephasing) and $p_v\le0.2$ (depolarizing) give $T\le-T_2\ln(1-2\cdot0.15)=\qnLifeVaultDeph\,T_2$ and
$T\le-\tau\ln(1-\tfrac43\cdot0.2)=\qnLifeVaultDep\,\tau$, rounded down. For the seal, every run used depolarizing noise
$p_0=0.005$, which attenuates each cell's Pauli eigenvalues by $1-4p_0/3$. Setting $e^{-T/\tau_s}=1-4p_0/3$ gives
$T\le\qnLifeSeal\,\tau_s$. A dephasing seal attenuates the $X$ and $Y$ components by $e^{-T/T_2}$ and leaves $Z$
unchanged, so matching the attenuation of the decaying components gives the same factor with $T_2$ in place of
$\tau_s$, and is conservative.

\emph{Bell-measurement errors.} For the pair state $|\Phi^+\rangle$, $(P\otimes Q)|\Phi^+\rangle\propto
(PQ^{\mathsf T}\otimes I)|\Phi^+\rangle$, so a Pauli error $P\otimes Q$ on the two halves changes the Bell outcome by
the product $PQ$ up to phase. Under two-qubit depolarizing noise of rate $p_{\mathrm{BM}}$ the 15 non-identity pairs
are equally likely. Three of them ($XX$, $YY$, $ZZ$) have an identity product, and the remaining twelve give $X$, $Y$
and $Z$ four times each. The outcome therefore suffers single-qubit depolarizing noise with non-identity probability
$\tfrac{12}{15}p_{\mathrm{BM}}=\tfrac45p_{\mathrm{BM}}$, which composes with the memory's noise on the vault half.

\emph{Frame.} A uniformly random Pauli per cell is two bits; a 32-cell module needs 64 bits, kept secret until
verification.